%% file: main.tex
\documentclass[sigconf,screen]{acmart}

\renewcommand\footnotetextcopyrightpermission[1]{} 

\usepackage{enumitem}
\usepackage[linesnumbered,ruled,lined,noend]{algorithm2e}
\usepackage[colorinlistoftodos,prependcaption,textsize=tiny]{todonotes}

\usepackage[utf8]{inputenc}
\usepackage{url}
\usepackage{tabularx}
\usepackage{amsmath,amsfonts}

\usepackage[noend]{algpseudocode}
\usepackage{amsthm}
\usepackage{booktabs}
\usepackage{multirow}
\usepackage{graphicx}
\usepackage{textcomp}
\usepackage{tikz}
\usepackage{xcolor}
\usepackage{enumitem}
 \usepackage{listings}

\usepackage{graphicx}

\usepackage{multirow}
\usepackage{listings}
\usepackage{color, soul}

\usepackage{textcomp}
\usepackage{framed}
\usepackage{hhline}
\usepackage{subcaption}
\usepackage[breakable]{tcolorbox}
\usepackage{float}
\usepackage{array}
\usepackage{makecell}
\usepackage{balance}
\usepackage{pifont}
\usepackage{xspace}
\usepackage{tcolorbox}
\usepackage{microtype}

\input{macro.tex}

\begin{document}

\title{Program Repair by Fuzzing over Patch and Input Space}

\input{authors.tex}
\input{abstract.tex}
\maketitle
\pagestyle{plain} 

\input{introduction.tex}
\input{motivation.tex}
\input{fuzz-repair.tex}
\input{cfr.tex}
\input{co-evolution.tex}
\input{evaluation.tex}
\input{related.tex}
\input{conclusion.tex}

\bibliographystyle{ACM-Reference-Format}
\bibliography{references}
\end{document}

%% file: macro.tex
\renewcommand{\shortauthors}{}

\newcommand{\tool}[1]{\textsc{#1}\xspace}

\newcommand{\toolName}{\tool{FuzzRepair}}

\newcommand{\genprog}{\tool{GenProg}}
\newcommand{\prophet}{\tool{Prophet}}

\newcommand{\cpr}{\tool{CPR}}
\newcommand{\senx}{\tool{Senx}}
\newcommand{\fixtofit}{\tool{Fix2Fit}}
\newcommand{\darjeeling}{\tool{Darjeeling}}

\newcommand{\vulnloc}{\tool{VulnLoc}}

\makeatletter
\def\BState{\State\hskip-\ALG@thistlm}
\makeatother

\algnewcommand\algorithmicforeach{\textbf{for each}}
\algdef{S}[FOR]{ForEach}[1]{\algorithmicforeach\ #1\ \algorithmicdo}

\definecolor{gray}{RGB}{215,215,215}
\sethlcolor{gray}

\newcommand{\ignore}[1]{}

\def\HiLi{\leavevmode\rlap{\hbox to \hsize{\color{gray!70}\leaders\hrule height .8\baselineskip depth .5ex\hfill}}}

\makeatletter
\newcommand{\removelatexerror}{\let\@latex@error\@gobble}
\makeatother

%% file: authors.tex
\author{Yuntong Zhang}
\email{yuntong@comp.nus.edu.sg}
\affiliation{%
  \institution{National University of Singapore\country{Singapore}}
}

\author{Ridwan Shariffdeen}
\email{ridwan@comp.nus.edu.sg}
\authornote{corresponding author}
\affiliation{%
  \institution{National University of Singapore\country{Singapore}}
}

\author{Gregory J. Duck}
\email{gregory@comp.nus.edu.sg}
\affiliation{%
  \institution{National University of Singapore\country{Singapore}}
}

\author{Jiaqi Tan}
\email{tjiaqi@dso.org.sg}
\affiliation{%
  \institution{DSO National Laboratories\country{Singapore}}
}

\author{Abhik Roychoudhury}
\email{abhik@comp.nus.edu.sg}
\affiliation{%
 \institution{National University of Singapore\country{Singapore}}
}

%% file: abstract.tex
\begin{abstract}
Fuzz testing (fuzzing) is a well-known method for exposing bugs/vulnerabilities in software systems.
Popular fuzzers, such as AFL, use a biased random search over the domain of program inputs, where 100s or 1000s of inputs (test cases) are executed per second in order to expose bugs.
If a bug is discovered, it can either be fixed manually by the developer or fixed automatically using an {\em Automated Program Repair} (APR) tool. Like fuzzing, many existing APR tools are search-based, but over the domain of patches rather than inputs.

In this paper, we propose search-based program repair as {\em patch-level fuzzing}.
The basic idea is to adapt a fuzzer (AFL) to fuzz over the patch space rather than the input space. Thus we use a patch-space fuzzer to explore a patch space, while using a traditional input level fuzzer to rule out patch candidates and help in patch selection. To improve the throughput, we propose a {\em compilation-free} patch validation methodology, where we execute the original (unpatched) program natively, then selectively {\em interpret} only the specific patched statements and expressions. Since this avoids (re)compilation, we show that compilation-free patch validation can achieve a similar throughput as input-level fuzzing (100s or 1000s of execs/sec). We show that patch-level fuzzing and input-level fuzzing can be combined, for a co-exploration of both spaces in order to find better quality patches. Such a collaboration between input-level fuzzing and patch-level fuzzing is then employed to search over candidate fix locations, as well as patch candidates in each fix location. 

Our results show that our tool FuzzRepair is more effective in patching security vulnerabilities than well-known existing repair tools GenProg/Darjeeling, Prophet and Concolic Program Repair (CPR). Moreover, our approach  produces other artifacts such as fix locations, and crashing tests (which show the evidence why patch candidates are ruled out). Thus our approach provides a pragmatic solution to enhance automation in program vulnerability repair, thereby reducing exposure of critical software systems to possible attacks. 
\end{abstract}

%% file: introduction.tex
\section{Introduction}

Software bugs are a perennial problem which incur significant economic costs. The 2020 Report from {\em Consortium for Information \& Software Quality} (CISQ) calculates the total cost of poor software quality in the United States to be \$2.08 trillion. 
Software debugging traditionally uses software developer manpower to (1) find the bug, (2) fix the bug, and (3) validate the correctness of the fix against the specification of the program. These activities are well known to be both challenging and time-consuming when performed manually.

Over the past decade, there has been significant progress in automated bug detection.
One popular technique is {\em fuzz testing}, which uses a (biased) random search over the space of program inputs (the input-space), typically testing 100s or 1000s of inputs per second.
The fuzzer may also use feedback from the program, such as branch coverage information, to bias the search into inputs which explore new paths in the program.
Fuzzing has been proven effective in real-world applications, with thousands of vulnerabilities discovered~\cite{fuzz-survey}.
Fuzz testing will report discovered bugs in the form of inputs that cause the program to crash or misbehave.
It is still up to the developer to patch the program accordingly, in order to resolve the bug.
Traditionally, debugging and patching is a manual effort.

One emerging alternative to manual debugging is \emph{Automated Program Repair} (APR)~\cite{LPR19}.
APR aims to {\em automatically} rectify software bugs without the need for developer intervention. Since APR promises to save both developer time and associated costs, it has received significant attention over the past decade, including the development of several tools and technologies. The general APR methodology works by automating various steps in the typical (manual) debugging workflow, including: {\em fix-localization} (i.e., {\em where} to apply the fix?), {\em patch-generation} (i.e., {\em how} to fix the bug?), and {\em patch-validation/ranking} (i.e., {\em validate} that the patch correctness).
Each of these sub-problems can be solved in different ways, leading to the development of many different APR tools and technologies, including
 \emph{semantic/constraint-solving based repair}~\cite{semfix},
 \emph{machine-learning}~\cite{prophet,bader2019getafix}
 \emph{templates}~\cite{tbar}, and
\emph{search}-based methods~\cite{genprog}, amongst others.

APR tools involve either an explicit or implicit navigation of the search space of program edits, as in generate-and-validate search-based repair tools. It is thus worthwhile to employ fast search space exploration tools to enable the search space navigation in automated program repair. A grey-box fuzzer is such a search exploration tool, an extremely effective one, which efficiently navigates the domain of program inputs. In this work, we re-purpose a grey-box fuzzer to work on the domain of program edits. However exploring the domain of program edits involves validating individual edits. This usually involves two steps (a) inserting the patch and re-compiling the program and (b) validating the patch against a test-suite. While re-purposing a grey-box fuzzer for navigating a space of program edits, we innovate along these two dimensions to achieve program repair via fuzzing. These innovations also seek to address two key challenges in program repair efficiency and effectiveness, namely : (a) recompilation affects program repair efficiency and (b) over-fitting patches affect program repair effectiveness. We now elaborate on these two points. 

One of the practical challenges in (the efficiency of) program repair is that patch validation can be a significant bottleneck.  For compiled programming languages, such as \verb+C+/\verb|C++|, the patch must be first \emph{applied} to the program and {\em recompiled} before it can be validated against a test suite. However, recompilation (including relinking) can be a relatively expensive operation, possibly in the order of seconds or minutes, depending on the size of the program.
This problem can severely limit both the {\em latency} (i.e., time to identify a plausible patch) and {\em throughput} (i.e., number of patches validated per time budget). This creates practical difficulties in real-life acceptance of program repair. The recent work of~\cite{trustAPR} shows that most real-world developers expect answers from APR tools in much shorter time frames, with 72\% of survey respondents preferring not even to wait longer than 30 minutes.
We therefore argue that the {\em latency} of repair tools is critically important, especially for real-world adoption, and is something that is largely neglected by most existing APR research.
To avoid recompilation of patch candidates, we propose to replace recompilation with a combination of {\em interpretation}
and {\em binary rewriting/probing}, in order to remove the {\em compiler-in-the-loop} from program repair---i.e., \emph{Compilation Free Repair} (CFR).
Compared to recompilation,
an interpreter can be low latency with a minimal startup time, allowing for patches to be validated immediately upon generation. Since whole-program interpretation is slow, our proposal uses {\em binary rewriting/probing} to limit the interpreted expressions/statements to those actually changed by the patch, leaving the rest of the program to use native (compiled) execution.
We show that compilation free repair can significantly improve the latency/throughput of repair tools, by order of magnitude.

Another practical challenge in (the effectiveness of) program repair comes from {\em patch over-fitting}. Even after a patch candidate is inserted and the program recompiled, the patched program is validated against a given test-suite. However a test-suite is an incomplete specification of program behavior. As a result, by searching and validating patch candidates, we may get patches which over-fit the given test-suite and may fail for tests outside the given test-suite. To ameliorate this challenge, we can embed a grey-box fuzzer working over program inputs into our patch-level fuzzing workflow. The input-level fuzzer will generate additional inputs and the patched program can be checked against these additional inputs against simple oracles such as crashes or hangs.  This will lead to a reduction in the patch pool, and the reduced patch pool can be used as seeds in the next iteration of patch level fuzzing. This leads to a fuzz campaign which iteratively alternates between patch-level fuzzing and input-level fuzzing until a time budget is exhausted. 

\paragraph*{Contributions:} The contributions of this paper can be summarized as follows:
\begin{itemize}[leftmargin=*]
    \item {\em Fuzzing as the search process in repair:} We observe that fuzzers represent an extremely optimized search process and as such can be repurposed for program repair. This is mostly a key implementation level observation. We note that genetic search has been widely used in generate and validate based repair. However, fuzzers represent an extremely optimized feedback driven search which we can exploit for repair. This implementation level observation allows us to achieve fast exploration of a large number of patch candidates. Indeed we conduct a search over the fix locations as well as the patch space at each fix location.
    \item {\em Compilation-free repair:} Since exploring large number of patch candidates involve recompilation, we develop compilation free repair as an enabling technology to achieve program repair via interpretation and binary rewriting. This leads to an order of magnitude improvement in repair latency / throughput, as we show with our experiments on known subjects in the security vulnerability repair benchmark \vulnloc~\cite{vulnloc}.
    \item {\em Reduce over-fitting for vulnerability repair:} The two contributions in the preceding allow for a test-based program vulnerability repair workflow via fuzzing. However to increase the effectiveness of repair and make the patches less overfitting, it is desirable to generate more tests. We integrate fuzzing based test generation into our program repair workflow, with the goal of ruling out over-fitting patches. We demonstrate its effectiveness specifically for security vulnerability repair on the \vulnloc benchmark \cite{vulnloc}, where it is found to be more effective in patching vulnerabities than existing program repair tools such as GenProg (and its new incarnation Darjeeling) \cite{genprog12}, Prophet \cite{prophet}, Fix2Fit \cite{fix2fit}, CPR \cite{cpr}, and \senx~\cite{senx}.
\end{itemize}

%% file: motivation.tex
\section{Motivation}

\definecolor{keyword}{rgb}{0.40, 0.40, 0}
\definecolor{comment}{rgb}{0, 0, 0.40}
\definecolor{const}{rgb}{0.40, 0, 0}
\definecolor{type}{rgb}{0, 0.40, 0}
\definecolor{typecheck}{rgb}{0.40, 0, 0.40}
\definecolor{row}{rgb}{0.7,0.95,0.95}
\definecolor{row2}{rgb}{0.95,0.95,0.95}
\definecolor{row3}{rgb}{0.90,0.90,0.90}
\definecolor{mygreen}{rgb}{0,0.5,0}
\definecolor{mygreen2}{rgb}{0,0.25,0}
\begin{figure}
\vspace{2mm}
{\scriptsize
\lstset{language=C,
                frame=single,
                numbers=left,
                numberstyle=\tiny,
                xleftmargin=2em,
                xrightmargin=4pt,
                numbersep=6pt,
                escapechar=@,
                basicstyle=\ttfamily\fontseries{m}\selectfont,
                deletekeywords={int},
                keywordstyle=\color{keyword}\ttfamily\fontseries{b}\selectfont,
                keywordstyle=[2]\color{type}\ttfamily\fontseries{b}\selectfont,
                keywordstyle=[3]\color{typecheck}\ttfamily\fontseries{b}\selectfont,
                keywordstyle=[4]\color{const}\ttfamily\fontseries{b}\selectfont,
                stringstyle=\color{red}\ttfamily,
                commentstyle=\color{comment}\ttfamily\fontseries{b}\selectfont,
                morekeywords=[2]{void, int, bool, int8_t},
                morekeywords=[3]{},
                morekeywords=[4]{true, false, NULL, 0},
                mathescape
}
\begin{lstlisting}[escapechar=@]
  bool bsearch(const int *a, int val, int lo, int hi) {
      while (lo <= hi) {
          // Bug: sub-expression (lo + hi) can overflow!
          int mid = (lo + hi) / 2;        
          if (a[mid] < val) lo = mid + 1;
          else if (a[mid] > val) hi = mid - 1;
          else return true; }
      return false;
  }
\end{lstlisting}
}
\caption{Implementation of sorted array membership using binary search.
This version contains an integer overflow bug shown in line 4.
\label{fig:example}
}
\end{figure}

\begin{example}[Buggy Binary Search]\label{ex:bsearch}
To illustrate program repair, we consider a buggy implementation of {\em binary search} algorithm as shown in Figure~\ref{fig:example}.
The algorithm searches for membership of a given value ($\mathtt{val}$) in a sorted array ($\mathit{a}$) within the range $\mathtt{lo}..\mathtt{hi}$.
The binary search algorithm works by repeatedly narrowing a range, until either (1) the matching value $\mathtt{val}$ is found (success), or (2) the range becomes empty (failure).
Each iteration of the binary search algorithm calculates the {\em midpoint} of the range using the statement $\mathtt{mid}{=}(\mathtt{lo}{+}\mathtt{hi}){\div}2$.
However, this version of binary search is famously vulnerable to an {\em integer overflow} bug that occurs when the sub-expression $(\mathtt{lo}{+}\mathtt{hi})$ exceeds the maximum integer value.
For illustrative purposes, we shall use 8 bit integers which overflow beyond the value (\verb+INT_MAX+).

The problem can be fixed by replacing the buggy line with an overflow-safe version, specifically with the patched assignment $\mathtt{mid}{=}\mathtt{lo}{+}(\mathtt{hi}{-}\mathtt{lo}){\div}2$.
Unlike the original statement, no sub-expression can cause an integer overflow, thereby resolving the bug.
$\qed$\end{example}

The goal of {\em Automated Program Repair} (APR) would be to automatically find this fix, based on a test suite provided by the user, and without any further intervention.
Search-based repair works by generating candidate patches, which are then validated against a suitable {\em test suite}.
For instance, a test-suite for Example~\ref{ex:bsearch} could include several test cases, where each test case is a call to $\mathtt{bsearch}$ and an expected result.
One example unit test for $\mathtt{bsearch}$ could be:
\begin{align*}\label{eq:test1}
    & \delta = \texttt{INT\_MAX} / 2 - 80\texttt{;} \\ 
    & \mathtt{assert}(\mathtt{bsearch}(\texttt{\{}1, 2, .., 100\texttt{\}}-\delta, 100, \delta, 99-\delta))\texttt{;} \tag{Test \#1}
\end{align*}
Here, the expected result is $\mathtt{true}$, i.e., the $\mathtt{bsearch}$ algorithm ought to find the element $100$ in the given sorted array of length $100$. 
However, the buggy implementation of $\mathtt{bsearch}$ will fail this test case.
During the first three iterations of the Figure~\ref{fig:example} algorithm, we have $(\mathtt{lo},\mathtt{hi}) = (0,99), (50,99), (75,99)$, relative to $\delta$, respectively.
However, during the fourth iteration that sub-expression $(\mathtt{lo}{+}\mathtt{hi})$ overflows, leading to a negative value for $\mathtt{mid}$ and the program crashing. (Search-based) APR tools will navigate over the space of the program edits or patches.
This will include a patch replacing the buggy line 3 with an integer overflow safe version, specifically $\mathtt{mid}{=}\mathtt{lo}{+}(\mathtt{hi}{-}\mathtt{lo}){\div}2$. However, the patch space is also very large, and includes many incorrect or irrelevant patches that do not fix the bug---or worse, introduce new bugs.
Program repair thus has well-known challenges, including  {\em performance}, {\em overfitting} and {\em localization} - which we discuss.

\paragraph{Performance}
Since the patch space can be very large, it may be necessary to validate thousands/millions of candidate patches before a plausible fix is found.
This challenge is exacerbated by the high costs of validating individual patches. To manually validate a patch $p$ for C/C++ programs, the software developer will first apply $p$ to the project's source file(s) and then {\em recompile} a patched variant of the program.
The problem is that recompilation (including re-linking) is a relatively costly operation, and may take in the order of seconds or minutes depending on the size of the project.

The recent work of~\cite{trustAPR} showed that most developers expect APR tools to provide answers promptly, with 72\% of survey respondents preferring to wait no longer than 30 minutes, with the unsurprising consensus being that faster is always better.
In contrast, most of the existing APR literature evaluates tools using a more generous fixed time budget, with 10/12/24 hours being typical.  The practical usage thus becomes limited --- essentially limiting APR to {\em offline} repair. We provide a {\em compilation free repair} approach based on binary rewriting which greatly speeds up searching over many patch candidates - since we do not need to recompile the program for every patch candidate.

\paragraph{Overfitting}

Another well-known challenge of APR is the {\em overfitting problem}.
This occurs when a candidate patch passes the (imperfect) oracle used for validation, such as a {\em test-suite}, but fails to generalize to other inputs.
For example, if we use
(\ref{eq:test1}) as a single-test test-suite, then the patch $\mathtt{mid}{=}\delta{+}99$ would be deemed {\em plausible} since the test-suite passes.
However, this patch merely fits to this specific test case (\ref{eq:test1}), and does not generalize to other test cases, and thus is an example of an {\em overfitting} patch.
The overfitting problem can be rectified using a stronger oracle, such as a more comprehensive test-suite.
However, manually writing test cases can be burdensome. This problem is even more pronounced for repairing security vulnerabilities where only a single failing test (the exploit) is available.
To help mitigate the problem, other oracles have been proposed, such as the {\em crash-freedom} oracle, for repairing security vulnerabilities. 
Crash-freedom can be used as a weak oracle, meaning that a test is deemed {\em passed} if the input does \textbf{not} lead to a crash. This allows us to generate additional tests (besides the exploit) using fuzzing.

\paragraph{Localization}  A final key challenge for program repair is the fix localization problem itself - deciding where in the program to generate the patch candidates. Many techniques use statistical fault localization as a proxy for fix localization though the two are different problems. Semantic repair tools have tried to combine localization and patch synthesis as a giant constraint solving problem achieved by a MaxSMT solver \cite{directfix}, but this has obvious negative repercussions on scalability. Our work repairs security vulnerabilities via a cooperation between fuzzers where the biased random search in the fuzzers searches over additional test inputs, fix locations, and patch candidates. This fast search is achieved by judicious use of fuzzers which represent a highly optimized search tool. 

Overall, program repair (APR) has been widely studied as a search-problem~\cite{genprog12,prophet} that navigates the space of program edits (a.k.a. {\em patches}) in order to find the {\em correct} program satisfying some fitness criteria, such as passing a test-suite.
Similarly, greybox fuzz testing tools, like AFL, use an evolutionary algorithm to navigate the {\em input} search space, using lightweight feedback (e.g., branch coverage) to guide the search toward interesting inputs that reveal the existence of program errors.
In this work, our aim is to adapt some of the success of fuzz testing into search-based program repair.
We observe that fuzz testing and repair essentially form a {\em duality}---i.e., fuzzing aims to {\em find} bugs, and program repair aims to {\em fix} bugs---and both are based on evolutionary algorithms.
Our underlying approach is therefore to treat search-based program repair as a form of fuzzing.
However, instead of fuzzing inputs to discover bugs, our approach is to fuzz {\em patches} to discover repairs---i.e., {\em patch-level fuzzing}, while input-level fuzzing starts with a program and attempts to find failing inputs, patch-level fuzzing starts with a buggy program and attempts to find {\em non-overfitting} patches as plausible repairs.
We now summarize the main design decisions.

\paragraph{Algorithm}
Our basic approach is to implement program repair as patch-level fuzzing.
For this we show that traditional fuzz testing tools, such as AFL~\cite{afl}, can be repurposed to fuzz over the space of patches rather than inputs.
The basic idea is to modify the fuzzer to maintain a queue of {\em interesting} patches rather than inputs, where the definition of ``interesting'' depends on a patch ranking heuristic.
Like traditional AFL, the main fuzzing loop periodically selects a patch from the queue for mutation.
Next, a set of mutant patches are generated which are then validated against the buggy program for {\em plausibility}---i.e., does the mutant patch pass the {\em oracle}?
Furthermore, {\em interesting} mutants may also be added to the queue for further mutation.
The output of the fuzzing process is a set of plausible patches, which can then be sorted by the patch ranking heuristic.
We next discuss how the performance and overfitting challenges are tackled in our approach.

\begin{figure*}[!t]
\centering
\includegraphics[scale=1.3]{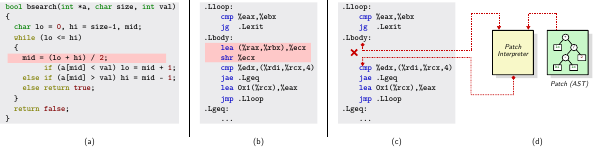}
\caption{An illustration of {\em patch interpretation}.
Here,
an implementation of binary-search which contains an integer overflow bug (a),
the program compiled into assembly (b).
The bug in (a) and (b) is highlighted.
Sub-figures (c) and (d) illustrate the implementation of {\em patch interpretation}.
Here, the instructions corresponding to the buggy line (c) are replaced with a call to the patch interpreter using binary rewriting.
The interpreter executes the replacement line (d) before returning control-flow back to the main program.\label{fig:interpret}}
\end{figure*}

\paragraph{Performance}
Input-level fuzzers, such as AFL, typically achieve throughputs of 100s or 1000s executions per second.
In contrast, the throughput of current generation of search-based tools can be orders of magnitude less.
As noted above, the underlying problem is that each mutant patch must be applied to the buggy program and {\em recompiled}, before the patch can be validated.
As (re)compilation is an expensive process, this quickly becomes a performance bottleneck.  Since (re)compilation is a major performance bottleneck in the current generation of search-based repair tools, we propose selective {\em patch interpretation} as an alternative.
We run the original (unpatched) executable natively, but also replace patch locations with a call to an {\em interpreter} that is injected into the program using {\em binary rewriting}.
The interpreter executes the patch statement(s), in place of the original code, before returning control-flow back to the native code.
Since most candidate patches only affect one (or a few) statements, a majority of the program still executes at native speed.
We shall refer to this method as {\em Compilation-Free Repair} (CFR).

\begin{example}
An example of {\em patch interpretation} is illustrated in Figure~\ref{fig:interpret}.
Here, we use the buggy version of {\em binary search} algorithm introduced in Example~\ref{ex:bsearch}.
This version contains a potential {\em integer overflow bug} with the statement $\mathtt{mid}{=}(\mathtt{lo}{+}\mathtt{hi}){\div}2$.
The problem can be fixed by replacing the buggy line with an overflow-safe version, specifically $\mathtt{mid}{=}\mathtt{lo}{+}(\mathtt{hi}{-}\mathtt{lo}){\div}2$.
Under {\rm patch interpretation}, we assume that the original (buggy) program has already been compiled into a binary executable $B$, as shown in Figure~\ref{fig:interpret}~(b).
Here, the instructions corresponding to the buggy line from Figure~\ref{fig:interpret}~(a) have also been highlighted.
Given a patch $P$, {\em patch interpretation} works by
\begin{enumerate}[leftmargin=*]
    \item {\em Diverting} control-flow to/from the patched program locations.
    In this example, the patch location is the highlighted buggy line from Figure~\ref{fig:interpret}~(a).
    \item {\em Interpreting} the patched statements/expressions, as illustrated in Figure~\ref{fig:interpret}~(d).
\end{enumerate}
An example of patch interpretation is illustrated in Figures~\ref{fig:interpret}~(c) and (d).
Here, the program will execute natively until the patch location is reached (at location \texttt{.Lbody}) corresponding to the buggy line in the original program.
Next, control-flow is diverted to a {\em patch interpreter}, which interprets the replacement line and updates the program state accordingly.
Finally, once the patch interpreter completes, control-flow is returned back to the original binary immediately after the patch location, and native execution resumes.
$\qed$\end{example}

\paragraph{Overfitting}
The problem of {\em overfitting} occurs when an incomplete oracle (e.g., a test-suite) is used as the basis for patch validation.
A candidate patch that passes the test suite will be deemed plausible if it passes all the tests in the test-suite, though it may fail other tests.  The problem can be solved using a {\em precise} oracle such as a formal specification. However, precise oracles are rarely available in practice. Instead, our solution is to automatically {\em extend} the existing imprecise oracle ``on-demand'', in the form of additional test cases. However, the expected output for a given generated test input is not known. These additional test cases (generated by fuzzing) will use {\em crash-freedom}~\cite{fix2fit} as the passing criteria---i.e., an extended test will be deemed {\em passed} if the input does not cause the program to crash.
The additional test-cases will be generated using input-level fuzzing, which co-operates with patch-level fuzzing as follows. 
\begin{enumerate}[leftmargin=*]
\item A {\em patch-level} fuzzer {\em adds} plausible patches to the pool that pass the current test oracle; and
\item An {\em input-level} fuzzer {\rm removes} overfitting patches from the pool that violate {\em crash-freedom} on a newly generated test case.
The shared test-suite is also updated with the new test case.
\end{enumerate}

Consider the overfitting patch ($\mathtt{mid}{=}\delta{+}99$) from above.
Assuming that the function parameters are read from input, the patch will very quickly lead to a crash for any array of length ${<}99$.
Such an overfitting patch will be quickly detected by input-level fuzzing and removed from the pool.
In contrast, suppose the ``correct'' patch $\mathtt{mid}{=}\mathtt{lo}{+}(\mathtt{hi}{-}\mathtt{lo}){\div}2$ was generated by the patch-level fuzzer.
The correct patch will pass the test-suite, and be added to the patch tool.
Overall, the patch-level and input-level fuzzers {\em co-evolve} the pool of plausible patches.

%% file: fuzz-repair.tex
\section{Program Repair as Patch-Level Fuzzing}
\label{section:repair-via-fuzzing}

\input{main_algo.tex}

We discuss how the search for patch candidates can be accomplished via a fuzzer. 

\subsection{Fuzzing Algorithm}

Algorithm~\ref{algo:patch-fuzzing} (Figure~\ref{fig:fuzzing}) describes the basic patch-level fuzzing algorithm.\footnote{Algorithm~\ref{algo:input-fuzzing} describes an input-level fuzzing algorithm for {\em co-evolution}, which will be detailed later in Section~\ref{section:co-evolution}.}
Here, the patch-fuzzer is provided with
a (buggy) program ($\mathit{Prog}$),
an initial set of {\em seed patches} ($\mathbb{P}_\mathit{seed}$),
a test oracle ($\mathbb{T}_\mathit{oracle}$), and some resource budget ($B$) such as time.
The output of the patch-fuzzer is a set of patches ($\mathbb{P}_\mathit{plausible}$) deemed plausible by the test oracle.
For our purposes, a \emph{patch} is a tuple $\langle\mathcal{L}, \mathit{Stmt}\rangle$, where $\mathcal{L}$ identifies some source location (e.g., file + line number), and $\mathit{Stmt}$ is a C/C++ \emph{statement} that replaces the original statement at location $\mathcal{L}$.
Algorithm~\ref{algo:patch-fuzzing} implements a basic fuzzing loop
that maintains a queue of patches ($\mathbb{P}_\mathit{queue}$) that is initialized to the initial seeds (line 1).
The main loop selects a patch $p$ from the queue ($\mathit{SelectNext}$, line 3), then generates a number of mutant patches $p'$ from $p$ ($\mathit{Mutate}$, line 5) controlled by a {\em power schedule} ($\mathit{energy}$, line 4).

Each generated mutant patch $p'$ is evaluated for fitness against two main criteria:
\begin{enumerate}[leftmargin=*]
\item $\mathit{IsPlausible}$, line 6: The mutant patch $p'$ is {\em applied} to buggy program $\mathit{Prog}$ to yield the mutant program $\mathit{Prog}'$.
The $\mathit{Prog}'$ program is then evaluated against the test oracle $\mathbb{T}_\mathit{oracle}$.
Here we assume that $\mathbb{T}_\mathit{oracle}$ is a test-suite with at least one failing test case.
Program $\mathit{Prog}'$ (and by extension $p'$) is deemed \emph{plausible} if all tests from $\mathbb{T}_\mathit{oracle}$ \emph{pass}.
\item $\mathit{IsInteresting}$, line 8: The mutant program is evaluated by some {\em interesting} metric.
The precise definition of {\em interesting} is flexible, but usually means that some new behaviour is observed by $\mathit{Prog}'$.
This can be
new program outputs, or new branch coverage observed, where the run-time observation (during the fuzz campaign) is aided by compile-time instrumentation.
\end{enumerate}
Plausible patches are saved into the $\mathbb{P}_\mathit{plausible}$ set, which is to be returned once the resource budget $B$ is reached, and interesting patches are added back to the queue $\mathbb{P}_\mathit{queue}$ for further fuzzing.

Algorithm~\ref{algo:patch-fuzzing} is similar to the search used by traditional input-level fuzzers, such as AFL~\cite{afl}.
The main differences are that (1) Algorithm~\ref{algo:patch-fuzzing} mutates \emph{patches} rather than \emph{inputs}, and (2) Algorithm~\ref{algo:patch-fuzzing} maintains a set of plausible patches.
Otherwise, the basic structure of the algorithms is similar.
Algorithm~\ref{algo:patch-fuzzing} is also similar to the genetic algorithms used by search-based repair tools, such as GenProg~\cite{genprog}, highlighting how fuzzing and repair algorithms are conceptually related.
We shall now describe each component of the algorithm in more detail.

\paragraph*{Seed Patches and Fix Localization}
Like input-level fuzzers, the patch-level fuzzing algorithm needs an initial set of {\em seed patches}.
Algorithm~\ref{algo:patch-fuzzing} accepts seed patches from any source, including user-suggested patches or those generated by other APR tools.
It is also possible to generate {\em default} seed patches for a given location $\mathcal{L}$, defined as $\langle \mathcal{L}, \mathit{Stmt}_\mathcal{L} \rangle$, where $\mathit{Stmt}_\mathcal{L}$ is the original statement at source location $\mathcal{L}$ in program $\mathit{Prog}$.
Algorithm~\ref{algo:patch-fuzzing} does not have an explicit {\em fix localization} step per se.
Instead, the initial set of fix location(s) is implied by the set of seed patches, and Algorithm~\ref{algo:patch-fuzzing} will search over the entire set if multiple locations are provided.
Over time, the fuzzing algorithm will naturally favor ``interesting'' locations with an observable effect on the test oracle $\mathbb{T}_\mathit{oracle}$. 
For our experiments (Section~\ref{sec:evaluation}), we consider {\em all} locations in the given exploit trace as possible fix locations, and Algorithm~\ref{algo:patch-fuzzing} is seeded with the corresponding default patch(es).

\begin{figure*}[t]
\centering
\includegraphics[scale=1.5]{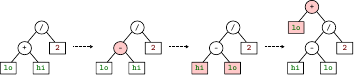}
\caption{An illustration of {\em patch mutation}.
Here, the original buggy expression $(\mathtt{lo}{+}\mathtt{hi}){\div}2$ is represented as an AST, and undergoes three mutations ({\em binary operator substitution}, {\em binary argument swap}, and {\em add term}) to derive the corrected expression $\mathtt{lo}{+}(\mathtt{hi}{-}\mathtt{lo}){\div}2$. \label{fig:mutations}}
\end{figure*}

\paragraph*{Patch Mutation}
The $\mathit{Mutate}$ operation takes a patch $p$ and applies a mutation to generate a new $p'$.
For this, we implement a set of standard {\em mutation operators} $m: \mathit{Stmt} \mapsto \mathit{Stmt}$, including \emph{Absolute Value Insertion} (ABS),
\emph{Operator Replacement} (OR),
\emph{Unary Operator Insertion/Deletion} (UOI/UOD),
\emph{Scalar Variable Replacement} (SVR), etc.
For our purposes, the $\mathit{Stmt}$ is an \emph{Abstract Syntax Tree} comprising terminal (e.g., variables, constants) and non-terminal nodes (e.g., operators).
Mutation operators are applicable to specific nodes in the AST.
By design, $\mathit{Mutate}$ does not change the patch location(s) $\mathcal{L}$, which is controlled by the initial set of seed patches (see above).
Mutations are applied according to a \emph{mutation schedule}.
Initially, a set of \emph{deterministic} mutations is tried, which
explores the space of ``simple'' fixes, such as all single-node mutations.
Next, any number of \emph{random} mutations will be tried, which includes multiple mutations over more than one node.
This design is analogous to input-level fuzzing tools, such as AFL~\cite{afl}, which tries {\em deterministic} (e.g. \texttt{bitflip}) before {\em random} (e.g. \texttt{havoc}) mutations.
An example of patch mutation is shown in Figure~\ref{fig:mutations}.
Here, we consider the buggy binary search algorithm from Example~\ref{ex:bsearch}.
The original buggy expression $(\mathtt{lo}{+}\mathtt{hi}){\div}2$ is represented as an AST in Figure~\ref{fig:mutations}~(a).
The corrected expression $\mathtt{lo}{+}(\mathtt{hi}{-}\mathtt{lo}){\div}2$ can be derived by the application of three mutation operators, as shown in Figure~\ref{fig:mutations}~(b), (c) and (d), and will be deemed {\em plausible} when validated against the test-suite $\mathbb{T}_\mathit{oracle}$.
Other patches can be generated by applying different mutations, however, most will fail $\mathbb{T}_\mathit{oracle}$.

\paragraph*{Plausible Patches and the Test Oracle}
We assume, as given, some initial test suite $\mathbb{T}_\mathit{oracle}$ that contains at least one failing test case.
A patched program $\mathit{Prog}'$ (and corresponding patch $p'$) is deemed {\em plausible} if $\mathit{Prog}'$ {\em passes} for each $t \in \mathbb{T}_\mathit{oracle}$.
Here, {\em pass} can mean that it produced the expected user-supplied output, or {\em crash-freedom}---i.e., the program does not crash when executed with $t$ as the input.
In principle, $\mathbb{T}_\mathit{oracle}$ can be large, meaning that it can be expensive to run the entire test suite for each generated patch candidate.
As an optimization, the $\mathit{IsPlausible}$ operation will prioritize a single failing test case  $t_\mathit{bad} \in \mathbb{T}_\mathit{oracle}$ to quickly filter out bad patches. If $\mathit{Prog}'$ fails with $t_\mathit{bad}$ as input, the corresponding patch is not plausible and no further testing is required.
Otherwise, the next failing (bad) tests are tried, followed by all non-failing tests, and the process will stop on the first failure.
If all tests pass, the patch is plausible, so $p'$ will be saved into $\mathbb{P}_\mathit{plausible}$.

\paragraph*{Interesting Patches}
Algorithm~\ref{algo:patch-fuzzing} requires some notion of {\em interesting} patches.
For (input-level) greybox fuzzing, this is usually  {\em code coverage}. since greater coverage corresponds to greater bug detection.
For patch-level fuzzing, we use the following heuristic:
\begin{enumerate}[leftmargin=*]
\item \emph{Plausible}: Plausible patches are generally interesting; or
\item \emph{Test Coverage}: The patch $p'$ passes a new failing test $t \in \mathbb{T}_\mathit{oracle}$ not previously passing; or
\item \emph{Branch Coverage}: The patch $p'$ passes the same tests as a previously queued patch, but exhibits different \emph{branch coverage}.
\end{enumerate}
The intuition is that any patch that passes a new test is making ``progress'' (increase in fitness), and thus can be queued.
Patches that do not pass new tests, but otherwise exhibit some new observable behaviour (such as increased branch coverage), may also be deemed interesting.
For this we reuse the existing branch coverage feedback using the standard AFL instrumentation.

\paragraph*{Implementation}
We have implemented a version of our design based on AFL~\cite{afl} (a famous input-level fuzzer).
The main changes to AFL include:
(1) implementing {\em patch mutation} ($\mathit{Mutate}$, line 5), 
(2) patch {\em validation} by applying and executing mutant versions of the program, and
(3) applying the {\em test schedule} to identify (un)interesting patches quickly.
To minimize the necessary changes, we also use a {\em flattened} patch representation that serializes the AST into flat binary files, which the AFL infrastructure can already manage.
Otherwise, the overall structure of AFL's main fuzzing loop remains intact.

%% file: main_algo.tex
\begin{figure*}
\SetKwIF{If}{ElseIf}{Else}{if}{}{else if}{else}{end if}%
\begin{minipage}{.46\textwidth}
\removelatexerror
\centering
\begin{algorithm}[H]
{
	\KwIn{Buggy program $\mathit{Prog}$, \\
              $\qquad$ test-suite $\mathbb{T}_\mathit{oracle}$, \\
              $\qquad$ seed patches $\mathbb{P}_\mathit{seed}$, \\
              $\qquad$ and resource budget $B$
	}
	\KwOut{A set of \emph{plausible patches}.}
        $\mathbb{P}_\mathit{plausible} \gets \emptyset$\texttt{;}\ \ $\mathbb{P}_\mathit{queue} \gets \mathbb{P}_\mathit{seed}$ \\
	\While{B}{
        $p \gets \mathit{SelectNext}(\mathbb{P}_\mathit{queue})$ \\
         \For{$i \in 1..\mathit{energy}(p)$}{
            $p' \gets \mathit{Mutate}(p)$ \\
            \If{$\mathit{IsPlausible}(\mathit{Prog}, p', \mathbb{T}_\mathit{oracle})$}{
               $\mathbb{P}_\mathit{plausible} ~\texttt{+=}~ \{p'\}$ \\      
            } 
             \If{$\mathit{IsInteresting}(\mathit{Prog}, p')$} {
                $\mathbb{P}_\mathit{queue} ~\texttt{+=}~ \{p'\}$  \\
            }
           }
         }
	\Return $\mathbb{P}_\mathit{plausible}$
	\caption{Patch-Level Fuzzing}
	\label{algo:patch-fuzzing}
}
\end{algorithm}
\end{minipage}
\hfill
\begin{minipage}{.47\textwidth}
\removelatexerror
\centering
\begin{algorithm}[H]
{
	\KwIn{Buggy program $\mathit{Prog}$, \\
              $\qquad$ plausible patches $\mathbb{P}_\mathit{plausible}$, \\
              $\qquad$ seed tests $\mathbb{T}_\mathit{seed}$, \\
              $\qquad$ and resource budget $B$
	}
	\KwOut{A set of \emph{counter-examples}.}
        $\mathbb{T}_\mathit{implausible} \gets \emptyset$\texttt{;}\ \ $\mathbb{T}_\mathit{queue} \gets \mathbb{T}_\mathit{seed}$ \\
	\While{B}{
        $t \gets \mathit{SelectNext}(\mathbb{T}_\mathit{queue})$ \\
         \For{$i \in 1..\mathit{energy}(t)$}{
            $t' \gets \mathit{Mutate}(t)$ \\
            \If{$\mathit{IsImplausible}(\mathit{Prog}, t', \mathbb{P}_\mathit{plausible})$}{
               $\mathbb{T}_\mathit{implausible} ~\texttt{+=}~ \{t'\}$ \\      
            } 
             \If{$\mathit{IsInteresting}(\mathit{Prog}, t')$} {
                $\mathbb{T}_\mathit{queue} ~\texttt{+=}~ \{t'\}$  \\
            }
           }
         }
	\Return $\mathbb{T}_\mathit{implausible}$
	\caption{Input-Level Fuzzing}
	\label{algo:input-fuzzing}
}
\end{algorithm}
\end{minipage}
\caption{Dual patch-level and input-level fuzzing.
The patch-level fuzzer can be used to generate more {\em plausible patches}, whereas the input-level fuzzer can be used to generate more {\em counter-examples} that refute plausible patches.
Both algorithms can be combined concurrently, to ``co-evolve'' the pool of plausible patches.
\label{fig:fuzzing}}
\end{figure*}

%% file: cfr.tex
\subsection{Compilation-Free Repair}
\label{section:cfr}

Algorithm~\ref{algo:patch-fuzzing} requires each patch candidate to be {\em applied} to the buggy program (see $\mathit{IsPlausible}$, line 6),
which is typically implemented by {\em recompilation} (this can be inefficient).
In this section, we introduce {\em Compilation-Free Repair} (CFR) as a method for validating the plausibility of candidate patches without the need to recompile a patched version of the buggy program.
CFR works by selective {\em patch interpretation}, meaning that the buggy program executes natively except for patched statements, which are executed using an interpreter.
CFR is illustrated in Figure~\ref{fig:interpret}.
The design of the patch interpreter is very similar to the built-in interpreter used by standard debugging tools, such as GDB~\cite{gdb}.
For example, at a given breakpoint, the GDB ($\texttt{print}~\mathit{expr}$) command will interpret the expression $\mathit{expr}$ with respect to the current program state, and will also update the program state if $\mathit{expr}$ has {\em side-effects} (e.g., $x\texttt{++}$, $x\texttt{=}y$, etc.).
The design of the patch interpreter for CFR is similar to that of the GDB expression interpreter, as discussed below.

\paragraph{Reading/Writing program state}
For our purposes, a {\em patch} $p$ is a pair comprising a {\em source location} $\mathcal{L}$ and an {\em Abstract Syntax Tree} (AST) representation of a statement $\mathit{Stmt}$, such as that illustrated by Figure~\ref{fig:mutations}.
The $\mathit{Stmt}$ may consist of {\em terminals} such as {\em variables} ($x$, $y$, etc.) and {\em constants} ($1$, $-1$, etc.), or {\em non-terminals} such as \verb+C+ {\em operators} (\texttt{+}, \texttt{-}, etc.) including {\em assignments} (\texttt{=}, \texttt{+=}, etc.).
Operationally, the statement ($\mathit{Stmt}$) will be executed in place of the original statement at location $\mathcal{L}$.
Typically, the statement ($\mathit{Stmt}$) will use one or more {\em program variables}, meaning that it is necessary for the patch interpreter to read-from or write-to the program state.
To find the locations of variables at runtime, the implementation of the patch interpreter assumes that the binary has been compiled with {\em debug information} enabled, i.e., using the \verb+-g+ compiler flag.
This will cause the compiler to emit DWARF debug information~\cite{dwarf} into the compiled binary, which encodes variable location information in the form of DWARF {\em expressions}~\cite{dwarf} as well as other useful information.
Thus, for a given source location $\mathcal{L}$, the set of program variables and corresponding DWARF expressions can be retrieved at runtime.
For a given variable, the patch interpreter will evaluate the corresponding DWARF expression, which yields the variable's location (e.g., register or stack frame).
For example, the DWARF debug information for the program in Figures~\ref{fig:interpret} (a) and (b) will encode the mapping between instruction addresses to the originating source line of code, as well as the mapping of source variables ($\mathtt{lo}$, $\mathtt{hi}$, $\mathtt{mid}$) and the corresponding locations at runtime (\verb+%rax+, \verb+%rbx+, \verb+%ecx+).
The patch interpreter can use this information to read-from or write-to these locations when evaluating the patch statement ($\mathit{Stmt}$).

\paragraph{Patch Evaluation}
The patch statement ($\mathit{Stmt}$) uses an {\em Abstract Syntax Tree} (AST) representation.
The patch interpreter itself is a basic recursive AST evaluator, either evaluating values (rvals) or locations (lvals), under the \verb+C+ semantics.
For {\em variables}, the corresponding location is calculated using the DWARF debug information, as explained above.

\paragraph{Injecting the patch interpreter}
In order to apply the patch $p$, the patch interpreter must be injected into the original unpatched binary at source location $\mathcal{L}$. 
The first step is to map $\mathcal{L}$ to the corresponding {\em instruction address} in the binary program.
For this, we use the DWARF debug information, which encodes the $\mathit{source} \mapsto \mathit{address}$ mapping amongst other useful information. Once the instruction address is known, the next step is to inject a {\em detour} to the patch interpreter.
For this, we use a simple form of binary rewriting, similar to how debuggers implement {\em breakpoints}.
The basic idea is to overwrite the instruction at the target address with a {\em software trap} (\texttt{int3}) instruction, which will generate a \verb+SIGTRAP+ signal if executed.
The \verb+SIGTRAP+ signal can be caught by a signal handler, and the {\em program state} (a.k.a., {\em context}) will also be saved and passed in as an argument.
The patch interpreter will then be invoked by the signal handler, and will interpret the patch statement ($\mathit{Stmt}$), modifying the program state accordingly.
Once the patch interpreter completes, control-flow is returned to the {\em next} source location after $\mathcal{L}$ in the program, by writing the corresponding instruction address directly to the {\em instruction pointer} register (\verb+%rip+).
In effect, any instructions corresponding to the original (unpatched) statement are skipped (not executed), and the patch statement ($\mathit{Stmt}$) is executed in its place.
The patch interpreter detour is illustrated in Figure~\ref{fig:interpret}~(b).

\paragraph*{Discussion and Limitations}
By using a patch interpreter, rather than a compiler, we avoid the costs associated with recompilation and relinking. Our experiments (Section~\ref{sec:evaluation}) show that with selective patch interpretation, patches validation can achieve throughputs similar to that of input-level fuzzing, which is over 100s executions per second.
Selective interpretation is only suitable for ``local'' patches that do not affect other statements.
However, most patch candidates generated by state-of-the-art program repair tools are local, and are therefore within the scope of patch interpretation.
Our current implementation of patch interpretation cannot handle statements that change the control-flow, such as statements of the form ($\mathtt{if}(\mathit{expr}) ...$).
This is partly due to the limitations of the DWARF information, which was primarily designed for debugging, and does not store information such as branch targets.
However, this limitation can be mostly mitigated by refactoring the program to extract conditionals as assignments, e.g., ($x{=}\mathit{expr}\texttt{;}~\mathtt{if}(x) ...$), thereby allowing the assignment to be patched.
We have implemented an automated code refactoring tool using the {\em LLVM Compiler Infrastructure}~\cite{llvm}.
Using this mitigation, our approach is applicable in general for program repair. Nevertheless, we use fuzzing to generate tests to rule out patch candidates. This makes our approach essentially applicable to vulnerability repair  (or crash repair), as we discuss in the next section.

%% file: co-evolution.tex
\section{Fuzzing based Co-Evolution}
\label{section:co-evolution}
Section~\ref{section:repair-via-fuzzing} presented a search-based program repair methodology based on patch-level fuzzing.
Section~\ref{section:cfr} optimizes the underlying approach with {\em Compilation-Free Repair}, allowing new patch candidates to be evaluated with high throughput.
Although our basic approach can efficiently navigate a large search-space of program edits, it still suffers from a fundamental problem in program repair, namely \textit{overfitting}.
This problem occurs when candidate patches pass the given (imperfect) test-suite, but fail to generalize to the implied specification of the program.

Our basic approach is to automatically extend the test suite, thereby allowing for overfitting patches to be detected and excluded.
To do so, we will use {\em input-level} fuzzing in an attempt to generate new {\em test cases} that refute any potentially overfitting patches that have been generated so far.
Essentially, we are using input-level fuzzing for its traditional role: finding bugs---but this time, finding bugs introduced by overfitting patches rather than bugs in the original program.
Next, we shall combine both patch-level and input-level fuzzing for a simultaneous exploration of both patch and input space, allowing for the set of plausible patches to {\em co-evolve}.

\begin{figure*}[t]
\centering
\includegraphics[scale=1.6]{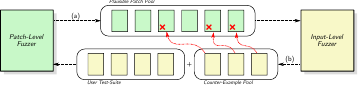}
\caption{An illustration of basic patch {\em co-evolution}.
In (a) the patch-level fuzzer adds new entries to the {\em plausible} patch pool, and in (b) the input-level fuzzers adds new entries to the {\em counter-example} pool.
For each counter-example, the set of plausible patches is also filtered.
At any given point, the patch-level fuzzer is guided by the test pool (user tests + counter-examples), and the input-level fuzzer is guided by the plausible patch pool.
\label{fig:co-evolve}}
\end{figure*}

\paragraph{Input-level fuzzing algorithm}
The input fuzzing algorithm is shown in Algorithm~\ref{algo:input-fuzzing} from Figure~\ref{fig:fuzzing}.
The algorithm structure is essentially the same as conventional input-level fuzzing, with a queue of {\em interesting} tests (initialized by $\mathbb{T}_\mathit{seed}$) and a main fuzzing-loop that repeatedly selects a queued test for mutation.
Each mutated test $t'$ is then validated against the program.
If the test $t'$ is deemed {\em interesting} (e.g., new branch coverage), it will be added to the queue for further mutation, and the process continues until some resource (e.g., time) budget $B$ is met.

In addition to the basic input-level fuzzing structure, Algorithm~\ref{algo:input-fuzzing} also checks each mutated test case $t'$ against a given set of {\em plausible patches} ($\mathbb{P}_\mathit{plausible}$).
The $\mathit{IsImplausible}$ test (line 6) 
holds if there exists a patch $p \in \mathbb{P}_\mathit{plausible}$ such that the corresponding
patched program $\mathit{Prog}'$ {\em fails} the mutant test $t'$.
Such a failing $t'$ is a witness to the {\em implausibility} of patch $p$, allowing $p$ to be excluded.
Furthermore, test $t'$ will be saved into the set $\mathbb{T}_\mathit{implausible}$, which can be used to extend the test oracle for patch-level fuzzing.
This will prevent $p$ (or similar patches) from being generated in the future.
In effect, Algorithm~\ref{algo:input-fuzzing} simultaneously refines the set of plausible patches as well as strengthens the test oracle.

We remark that Algorithm~\ref{algo:input-fuzzing} is the dual of Algorithm~\ref{algo:patch-fuzzing} under the syntactic substitution:
$$\{\mathbb{P} \mapsto \mathbb{T},
\mathit{plausible} \mapsto \mathit{implausible},
\mathit{IsPlausible} \mapsto \mathit{IsImplausible}\}$$
Essentially, the patch-level fuzzer is a process for {\em generating} plausible patches, and the input-level fuzzer is the dual process for {\em refuting} overfitting patches.
When combined, the two processes {\em co-evolve} the set of plausible patches.
We elaborate on this idea below.

\paragraph{Patch Co-evolution via Fuzzing}
We formulate the program repair problem as a co-operation between two fuzzers
which add/remove patches and tests to/from a common pool.
The Patch-Level fuzzer (Algorithm~\ref{algo:patch-fuzzing}) generates plausible patches that pass the (evolving) test-suite, whereas the Input-Level fuzzer (Algorithm~\ref{algo:input-fuzzing}) generates test-cases that remove over-fitting patches from an (evolving) patch pool.
Such a co-evolution, with two fuzzers, attempts to generate and refine the pool of patches.
An illustration of patch co-evolution is shown in Figure~\ref{fig:co-evolve}.

Each fuzzing process is also {\em dependent} on the other.
Specifically, the patch-level fuzzer evaluates patch candidates against the current {\em test-suite} ($\mathbb{T}_\mathit{oracle}$), which includes the initial seed tests (provided by the user) and any counter-example(s) generated by the input-level fuzzer.
Similarly, the input-level fuzzer evaluates tests against the current pool of plausible {\em patches} ($\mathbb{P}_\mathit{plausible}$).
The patch-level and input-level fuzzers can run concurrently using a simple time-sharing algorithm.
Essentially, at any given point in time, we define a numerical value $\mathit{Target}$ to be the desired number of patches in the plausible patch pool.
This leads to the following co-evolution algorithm illustrated by Figure~\ref{fig:co-evolve}:
\begin{enumerate}[leftmargin=*]
\item A {\em patch pool} $\mathbb{P}_\mathit{plausible}{=}\emptyset$ and a {\em test pool} $\mathbb{T}_\mathit{oracle}{=}\mathbb{T}_\mathit{user}$ is initialized, where $\mathbb{T}_\mathit{user}$ is the initial user-supplied test suite.
\item While $|\mathbb{P}_\mathit{plausible}|{<}\mathit{Target}$, we run the patch-level fuzzer (Algorithm~\ref{algo:patch-fuzzing}) to generate new patches to be added to the pool.
\item While $|\mathbb{P}_\mathit{plausible}|{\geq}\mathit{Target}$, we run the input-level fuzzer (Algorithm~\ref{algo:input-fuzzing}) to generate new counter-examples (added to $\mathbb{T}_\mathit{oracle}$).
For new counter-examples generated, the patch pool is filtered accordingly.
\end{enumerate}
Here, $\mathit{Target}$ is implementation-defined, and can be a dynamic value.
For example, if either fuzzer is running too long then the target can be adjusted accordingly.

\paragraph{Overfitting Detection and Patch Ranking}
Thus far, we have not defined the notion of {\em failure} for automatically generated test cases.
For this, there are two main possibilities: {\em crash-freedom} and {\em differential-testing}. Here, a patch $p$ passes the newly generated test $t$ if it does not cause the (patched) program to crash, which is a {\em hard} filter for patch implausibility. Crash-freedom is also enhanced by the use of {\em sanitizers}, specifically AddressSanitizer (ASAN)~\cite{asan} and {\em Undefined Behaviour Sanitizer} (UBSan)~\cite{ubsan}. If a patch $p$ is deemed crash-free, then the output of the patched program is compared with the original (buggy) program to detect differences. The intuition is that for non-crashing tests, the buggy and patched program should behave similarly, while for crashing tests, the buggy and patched program should behave differently~\cite{yu2017}. For example, given a divide-by-zero bug $x{=}y{\div}z$, any patch that fixes $z$ to a non-zero value will be deemed plausible but probably overfitting. Such overfitting patches may manifest other kinds of misbehaviour besides a crash, such as changing the program output. Finally, we perform {\em patch-ranking} using a {\em ranking function}, i.e., $\mathit{rank}(p)$ for the remaining (survived) patches that are deemed as ``passing'' by the evolved test-suite. Different patch-ranking heuristics can be used, and many have been proposed in APR literature~\cite{Xiong2018_PatchCorrectness}. For our implementation, we use a form of {\em differential testing} that compares the control-flow of the original (buggy) program and the patched version, with smaller differences receiving higher ranks.

\paragraph{Discussion and Limitations}
Combining patch-level and input-level fuzzing is a natural mitigation to the overfitting problem.
That said, crash-freedom is not always applicable, and differential-testing are not guaranteed to be accurate, meaning that the resulting patches may still be overfitting. 
Without a precise specification of the intended behaviour, the overfitting problem is inherent and cannot be completely eliminated.

%% file: evaluation.tex
\section{Evaluation}\label{sec:evaluation}

In this section, we present the experimental evaluation of our program repair technique. 

\subsection{Setup}
The goal of our work is to generate patches by exploring the patch-space of a buggy program using re-purposed fuzzing (see Section~\ref{section:repair-via-fuzzing}). We evaluate our implemented fuzzing-based program repair technique \toolName in fixing software vulnerabilities. The dataset we use is the \vulnloc benchmark~\cite{vulnloc} which consists of real-world C/C++ applications with a failing test-case that exposes a security vulnerability. There are 11 subjects with lines of code ranging from 8K to 2.7M, which consist of popular libraries such as LibJPEG and utilities such as Binutils. Vulnerabilities reported in six classes inclusive of buffer overflow, use-after-free, integer-overflow, null-pointer-deference, data-type overflow and divide by zero errors.

Our implementation of the fuzz search engine for repair is an extension of AFL~\cite{afl}. All experiments are conducted using Docker containers on top of AWS (Amazon Web Services) EC2 instances. We used the c5a.8xlarge instance type which provides 32 vCPU processing power and 64GB memory capacity. All experiments have been executed with a timeout of 1 hour. As confirmed by developer surveys, 1 hour is a reasonable expectation from developers for an automated technique to produce a patch ~\cite{trustAPR}. We note that many program repair works use higher timeouts such as 12 hours or 24 hours, which may not match developers' expectations.

\subsection{Fixing Vulnerabilities}
We evaluate the effectiveness of our technique \toolName in fixing security vulnerabilities in real-world applications. 
We compare the performance of \toolName with the state-of-the-art program repair tools for C/C++ programs to generate a plausible repair for the vulnerabilities in \vulnloc benchmark~\cite{vulnloc}. For this purpose, we select \prophet~\cite{prophet}, \darjeeling(\genprog)~\cite{genprog}, \cpr~\cite{cpr}, \senx~\cite{senx} and \fixtofit~\cite{fix2fit}. \prophet~\cite{prophet} is a learning-based repair technique that uses a correctness model to prioritize patch exploration and rank candidate patches. \darjeeling~\cite{darjeeling} mutates program statements using mutation operators extended from \genprog~\cite{genprog}. \cpr~\cite{cpr} is a semantic-based repair technique that uses program synthesis to generate patches. \senx is a vulnerability repair tool that generates patches using vulnerability-specific and human-specified safety properties. \fixtofit~\cite{fix2fit} is a search-based repair technique that implements an efficient search exploration strategy to navigate the patch-space. Table~\ref{table:comparison} shows the results of program repair tools for the security vulnerability benchmark \vulnloc~\cite{vulnloc}. Column $\#Vul$ reports the total number of vulnerabilities per subject program, which is 43. The rest of the columns indicate the number of bugs where a plausible patch was found for each tool. 
The two vulnerabilities in FFmpeg could not be reproduced in our experimental environment, and thus are left out for all the experiments.

\input{tab-comparison.tex}

Overall we find that \prophet and \darjeeling have the lowest count of bugs in which it was able to generate a plausible patch. We attribute this lower count to the constraint enforced on the time budget (i.e., 1 hour), which may affect the search process. One of the major bottlenecks in these two tools is the significant recompilation cost that prevents them from finding a plausible patch within the given time budget, as reported in previous studies as well~\cite{trustAPR}. \senx symbolically extracts the memory range access by a loop by performing loop cloning and access range analysis. However, loop cloning fails in many of the instances; hence it can only fix 13 bugs in \vulnloc benchmark. \fixtofit and \cpr perform reasonably well, with 33 and 35 bugs finding a plausible patch, respectively. \fixtofit enumerates the patch-space using an efficient exploration strategy which uses test-equivalence relations~\cite{f1x}, while \cpr explores its search-space using abstract patches~\cite{cpr}. \fixtofit extends the super-mutant generation for validation implemented in F1X~\cite{f1x}, which enables it to enumerate a large patch-space to find a plausible patch that can fix the vulnerability (i.e., a single failing test-case). \cpr uses concolic execution to reason about the candidate patches, which does not require executing the program multiple times. However, concolic execution itself is expensive. \cpr also requires the fix location to be provided so that its patch-space is restricted to a user-provided fix-location and user-configurable set of patch-ingredients. While it is effectively finding a plausible patch for 35 bugs, the search-space is considerably limited. In comparison, \toolName is able to outperform state-of-the-art tools by a significant margin, generating a plausible fix for 41 bugs. 

Our approach does not require the fix location to be provided. This is not surprising given the efficacy of fuzzing in efficiently exploring large search spaces. Thus we are benefiting from using fuzzing, a well-known optimized search process, as the core program repair technique. We note that even though other program repair works have used fuzzing as a helper technique such as  \cite{fix2fit} using fuzzing to generate additional test cases (to filter out patch candidates) --- our work is the first to conduct program repair itself (the patch space navigation) by fuzzing.  One key distinction between \toolName and \fixtofit is the instrumentation of the target program. Although both \toolName and \fixtofit use AFL~\cite{afl} as a back-end to perform fuzzing for its automatic test-generation, \fixtofit relies on compile-time instrumentation, while \toolName performs binary-rewriting~\cite{e9patch} to insert AFL instrumentation.

\begin{tcolorbox}[boxrule=1pt,left=1pt,right=1pt,top=1pt,bottom=1pt]
\textbf{Fixing Vulnerabilities:}
Experimental results shows that \toolName outperforms existing repair techniques in finding a plausible patch with a significant margin on the \vulnloc benchmark. 
\end{tcolorbox}

\subsection{Generating Patch-related Recommendations}
In addition to finding a plausible patch that fixes a vulnerability in software, \toolName can aid the developer by providing recommendations in terms of suggestions  to find a better fix. Developers may not always select the top-ranked patch that an automated patch generation tool produces. Providing additional insights on different fix-locations, recommendations and artifacts to fix the vulnerability can be helpful, as confirmed by developer surveys \cite{trustAPR}. In this section, we evaluate how \toolName can provide such additional insights and artifacts for the developer to find alternative fixes. 
Table~\ref{table:artifacts} shows the artifacts generated by \toolName while finding a fix for programs in \vulnloc benchmark~\cite{vulnloc}.  Given the pool of plausible patches generated by \toolName, the co-evolution process (refer to Section~\ref{section:co-evolution}) generates more test-cases to prune the overfitting patches according to a test oracle.
Column $P_{o,c}$ indicates the number of patches deemed overfitting by the oracle of crash-freedom. In addition, we used differential testing to identify overfitting patches. The assumption is that for any non-crashing test-case on the original buggy program, the patched program should have the same behavior. Column $P_{o,d}$ indicates the additional number of patches deemed overfitting by differential testing. For fix-localization, column $L_{total}$ indicates the total number of locations being considered by \toolName, and column $L_{rank}$ shows the rank of the location where developer's patch is, among all locations considered. We also report the number of test-cases generated from co-exploration, where $T_c$ and $T_p$ show the total number of crashing and non-crashing test-cases generated by the input-fuzzer, respectively. These test-cases are filtered out to exercise the fix-locations identified by the plausible patches in the patch pool, which serves as evidence for correct/incorrect behavior of the patches. 

\input{tab-artifact-summary.tex}

\subsubsection{Fix Locations}
\toolName combines fix-localization as part of the patch-generation process, where a location in the program is identified as a fix-location if a plausible patch can be generated. \toolName can determine a fix-location by quickly generating and validating a plausible patch. 
Table~\ref{table:artifacts} column $L_{total}$ indicates the total number of locations being considered by \toolName for a given bug, and column $L_{rank}$ shows the rank of the location where developer's patch is, among all locations considered.
On average \toolName finds the developer fix-locations ranked in top-5 for 17 bugs, and top-10 for 25 bugs in the \vulnloc benchmark. Traditional program repair techniques compute fix-locations using spectrum-based fault localization (SBFL) techniques to identify and rank suspicious (high probability of being the root cause of the bug) program locations.
If the correct location is not among the localized fix locations, the repair process will not be able to generate the correct patch~\cite{liu2019}. 
Thus, repair techniques would need to iterate over a list of possible fix-locations before finding a location that can generate a plausible patch. Due to the high patch validation cost resulting from re-compilation, it takes significant time to find a fix-location that can generate a plausible patch. In contrast, \toolName uses {\em compilation-free repair} that allows us to quickly generate and validate patches at a speed of 100s of patches per second. Hence, \toolName can quickly provide insights into fix-locations where plausible patches can be generated.

\begin{tcolorbox}[boxrule=1pt,left=1pt,right=1pt,top=1pt,bottom=1pt]
\textbf{Fix-Localization:}
\toolName can identify the developer fix-location in top-5 ranking for 17 instances in the \vulnloc benchmark which only provides one failing test-case.  
\end{tcolorbox}

\subsubsection{Over-fitting Patches}
One overlooked aspect of automated program repair is the insights provided by over-fitting patches. Overfitting patches fix the program for a given test-suite (for vulnerability repair, it is a single failing test or exploit) but fails on additional test-cases. Over-fitting patches can still convey valuable information about the fix-location. Such information can be exploited by human developers (or other specialized techniques) to generate better patches (e.g., see \cite{dirk2021} which refers to ``partial fixes''). \toolName generates a list of overfitting patches that fix the vulnerability but fails to generalize on additional test-cases. Shown in Table~\ref{table:artifacts} columns $P_{o,c}$ and $P_{o,d}$ are the total number of over-fitting fixes generated by \toolName. For each overfitting patch, our technique also generates information about which test case $t$ deemed it as an incomplete/incorrect patch. In fact, Table~\ref{table:artifacts} also reports the number of crashing and non-crashing test cases generated by the input-level fuzzer. 
On average, \toolName generates 777 and 2237 overfitting patches that failed to avoid the program crash and failed on differential testing, respectively. Note that the vulnerability in the given exploit is fixed, but the fixed program may still be crashing on other tests.  These overfitting fix candidates are internally used by the patch-fuzzer to generate better patches. Thus, given the fact that we use a fuzzer to mutate patch candidates --- overfitting patch candidates can be useful as the fuzzer can try to evolve them.  The intuition is that by mutating overfitting patches we can explore and discover ``better'' patches.

\begin{figure*}[t!]
    \centering
    \begin{subfigure}[]{0.3\textwidth}
        \centering
        \includegraphics[width=\textwidth]{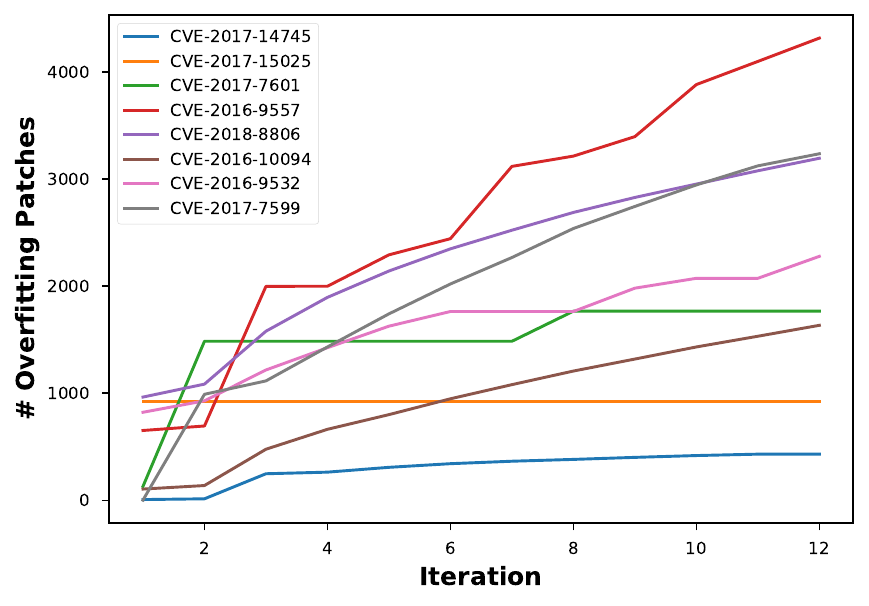}
        \caption{Discarding overfitting patches}
        \label{figure:patch-pruning}
    \end{subfigure}%
    ~ 
    \begin{subfigure}[]{0.3\textwidth}
        \centering
        \includegraphics[width=\textwidth]{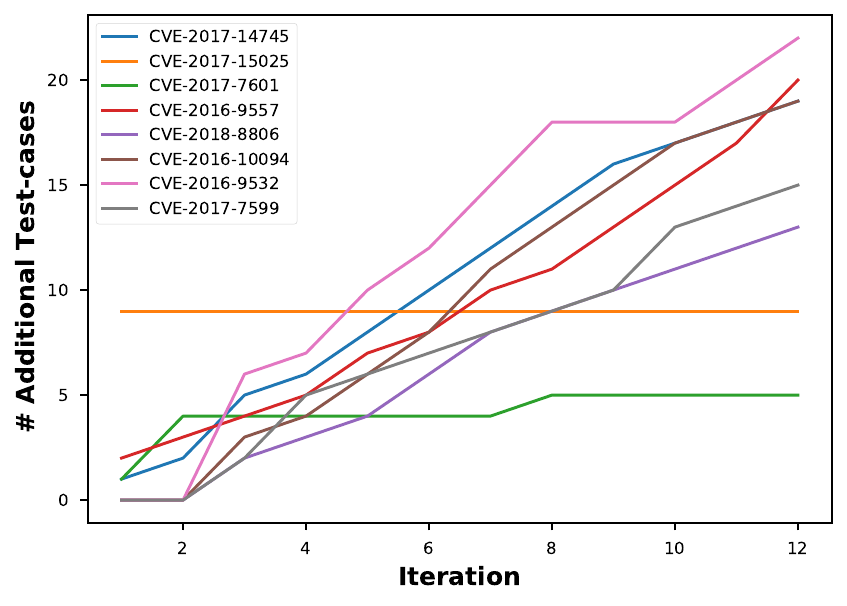}
        \caption{Test-suite enhancement}
        \label{figure:test-enhancement}
    \end{subfigure} %
   ~ 
    \begin{subfigure}[]{0.3\textwidth}
        \centering
        \includegraphics[width=\textwidth]{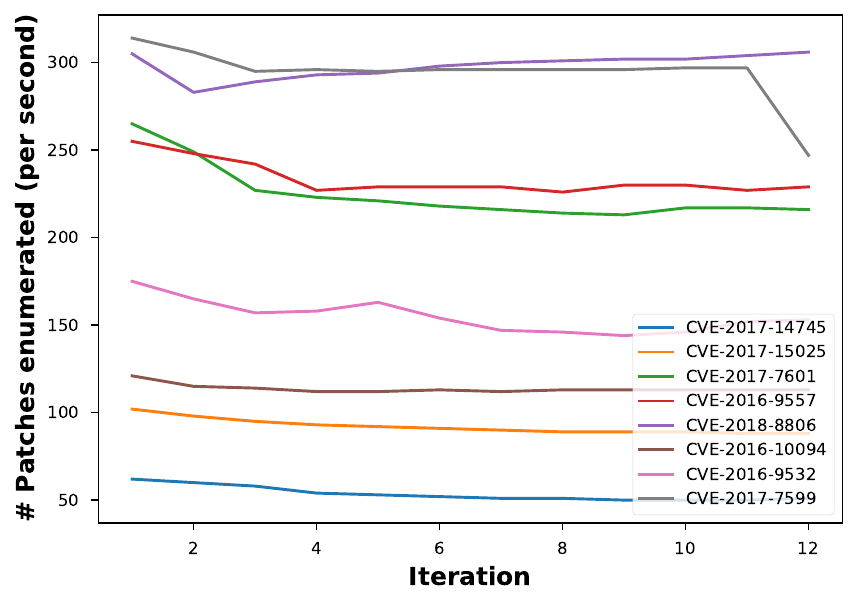}
        \caption{Patch-level fuzzing speed}
        \label{figure:fuzz-speed}
    \end{subfigure}
    ~ 
    \caption{Impact of Co-Evolution in \toolName}
    \label{figure:impact-of-co-evolution}
\end{figure*}

\subsubsection{Additional Test Cases}
In addition to finding fix-locations and over-fitting fixes as suggestions for the developer, \toolName can also strengthen the test-suite for the developer by generating new test-cases. \toolName is able to quickly generate crashing and non-crashing test-cases that exercise the patch locations considered by the patches in the patch-pool. In Table~\ref{table:artifacts},  columns $T_c$ and $T_p$ shows the total number of crashing and non-crashing test-cases generated for co-evolution by the input-fuzzer. On average, \toolName generates 579 new crashing test-cases and 146 non-crashing test-cases that can be used to identify over-fitting patches. These additional test-cases are internally used in \toolName by the patch-fuzzer and input-fuzzer. The additional test-cases prevent the patch-fuzzer from generating certain over-fitting patches. These test-cases also guide the input-fuzzer to find other ``interesting'' test-cases (since input-fuzzer can mutate them further) to refine the pool of patch candidates. 

\begin{tcolorbox}[boxrule=1pt,left=1pt,right=1pt,top=1pt,bottom=1pt]
\textbf{Improving Program Specification:}
\toolName generates new test-cases that exercise patch-locations to strengthen the specification for program repair for the \vulnloc benchmark. Overfitting patches are evolved by the patch level fuzzer via mutations.
\end{tcolorbox}

\subsection{Analysis of Co-Evolution}

In this section, we evaluate the impact of co-evolution in \toolName in terms of overfitting patch detection, test-suite enhancement, fuzzing speed, and ranking of the fix-locations. Figure~\ref{figure:impact-of-co-evolution} depicts the progress over each iteration, for the number over-fitting patches removed, the number of test-cases added to the test-suite, and the execution speed of the patch-level fuzzer.

For brevity, in Figure~\ref{figure:impact-of-co-evolution}, we select experiments that undergo a minimum of 12 iterations of co-evolution between patch-fuzzer and input-fuzzer. These experiments provide sufficient data to analyze the impact of co-evolution. Figure~\ref{figure:patch-pruning} shows the number of patches identified and discarded as over-fitting during co-evolution. The general trend is increasing, suggesting that with each iteration, \toolName identifies more over-fitting patches and discards them from the patch pool. Figure~\ref{figure:test-enhancement} illustrates the enhancement of the test-suite, where new test-cases that can identify over-fitting patches are generated and added to the existing test-suite. These new test-cases prevent similar patches from being generated in future iterations. In each iteration, the input-fuzzer focuses on the remaining patches in the patch-pool to find  inputs that can detect over-fitting behavior. In addition, we also analyzed the performance of our patch-level fuzzer, since the number of test-cases in the test-suite increases over time, which may require patch-level fuzzer to spend more time validating each candidate patch. Figure~\ref{figure:fuzz-speed} shows the speed of the patch-level fuzzer as the number of patches enumerated per second. In general, adding new test-cases does not have a significant effect on the patch-level fuzzing speed. Since fuzzing can efficiently evaluate 100s of test-cases per second, adding a few test-cases does not have a significant impact. CVE-2017-15025 is an outlier in Figure~\ref{figure:impact-of-co-evolution}, which does not find over-fitting patches since iteration \#2. This is because the input-level fuzzer did not generate new inputs that can differentiate existing patches. Hence the overall impact of co-evolution for \verb+CVE-2017-15025+ is not significant. We  strictly limit the time of individual test executions (default 20ms in AFL). This may drive the fuzzing search away from neighborhoods involving very large-sized inputs (such as large files), as is the case for this subject.  The issue can thus be potentially ameliorated with different timeout settings for test executions in the fuzzer.

Overall, considering all patch executions in all iterations, we have observed an average of 240 patches enumerated per second during experiments with \toolName. This patch execution speed is significantly faster than state-of-the-art search-based repair tools. For example, on average, Prophet \cite{prophet} validates seven patch templates per second, and Darjeeling \cite{darjeeling} validates six patches per minute based on our experimentation on the \vulnloc benchmark. The efficiency in patch validation in our tool is made possible by the enabling technology of compilation free repair.

\begin{figure}[t!]
    \centering
    \includegraphics[width=0.45\textwidth]{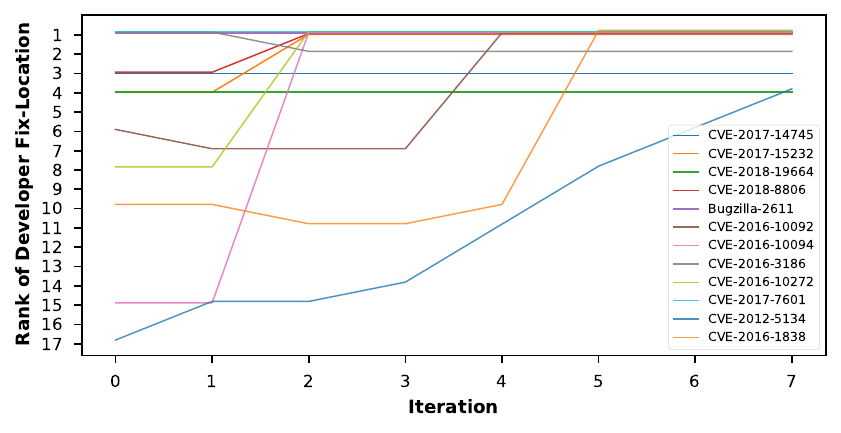}
    \caption{Impact of co-evolution on fix-location ranking. For each iteration, the data point represents the rank for the developer fix-location.}
    \label{figure:fix-loc-rank}
\end{figure}

Figure~\ref{figure:fix-loc-rank} shows the ranking of the developer fix-location. In 25 (out of 43) experiments, the developer fix-location was placed in top-10. Here, we analyze the placement of the developer fix-location in each iteration to understand the effect of co-evolution on the fix-location ranking. For this purpose, we select experiments with a minimum of 8 iterations of co-evolution and where the developer fix-location was found in top-5 listing. Figure~\ref{figure:fix-loc-rank} illustrates the rank of the developer fix-location from iterations 1 to 8. The results show that, with each iteration, the rank of the developer fix-location improves and in many cases ranked as high as top-1. 
During the co-evolution process, \toolName ranks the available fix-locations by the total number of plausible patches found at each location.
This improvement of fix-location ranking suggests that the co-evolution process generates more plausible patches at the correct location, potentially due to the stronger specification enforced by the evolving test-suite.
\verb+CVE-2012-5134+ is an example of such improvement where with each iteration, the rank is gradually improved from \#17 to \#4.

\begin{tcolorbox}[boxrule=1pt,left=1pt,right=1pt,top=1pt,bottom=1pt]
\textbf{Impact of co-evolution:}
Empirical evidence shows that co-evolution greatly improves the quality of the patches, and ranks the correct fix-location higher with less performance over-head. 
\end{tcolorbox}

%% file: tab-comparison.tex
\begin{table}[]
\centering
\caption{Comparison with program repair tools. The experiments have been executed with timeout of 1 hour.}
\footnotesize
\setlength{\tabcolsep}{3pt}
\begin{tabular}{|lr|rrrrrr|}
\hline
\textbf{Program} &
\textbf{\#Vul} &

\toolName & \prophet & \darjeeling  & \fixtofit & \cpr & \senx \\

\hline
\hline

Libtiff & 15 & 15 & 8 & 4 & 14 & 13 & 8  \\
Binutils & 4 & 4 & 0 & 0 & 1 & 3 & 0  \\
Libxml2 & 4 & 4 & 0 & 0 & 4 & 4  & 1  \\
Libjpeg & 4 & 4 & 2 & 2 & 3 & 4  & 1 \\
FFmpeg & 2 & - & - & - & - & - & -  \\
Jasper & 2 & 2 & 1 & 2 & 2 & 2 & 0  \\
Coreutils & 4 & 4 & 0 & 2 & 4 & 4 & 0 \\
LibMing & 3 & 3 & 0 & 1 & 3 & 1 & 1  \\
ZzipLib & 3 & 3 & 0 & 2 & 3 & 2 & 0 \\
LibArchive & 1 & 1 & 0 & 1 & 1 & 1 & 1  \\
Potrace & 1 & 1 & 0 & 0 & 1 & 1  & 1 \\

\hline
\hline

\textbf{Total} & 43 & 41 & 11 & 14 & 33 & 35 & 13 \\

\hline

\end{tabular}
\label{table:comparison}
\end{table} 

%% file: tab-artifact-summary.tex
\begin{table}[t]
\centering
\caption{Supplementary artifacts generated by \toolName on \vulnloc benchmark.}
\label{table:artifacts}
\footnotesize
\setlength{\tabcolsep}{3pt}
\begin{tabular}{|r|ll|cc|cc|cc|}
\hline
\textbf{ID} & 
{\textbf{Project}}  &{\textbf{Bug ID}} & 
 \textbf{$P_{o,c}$} & \textbf{$P_{o,d}$}  & \textbf{$L_{total}$} & \textbf{$L_{rank}$} &
\textbf{$T_c$} & \textbf{$T_p$}  \\

\hline
\hline

1  & binutils  & CVE-2017-14745  & 538  & 0  & 12  & 4  & 817  & 15  \\
2  & binutils  & CVE-2017-15020  & 2317 & 0  & 24  & 6  & 1822 & 25  \\
3  & binutils  & CVE-2017-15025  & 670  & 906 & 406  & 17 & 168  & 117 \\
4  & binutils  & CVE-2017-6965   & 4    & 536  & 349  & -  & 1172 & 163  \\
5  & coreutils  & gnubug-19784   & 0  & 36549  & 26  & 2  & 8  & 5   \\
6  & coreutils  & gnubug-25003   & 0  & 22460  & 72  & 14  & 1  & 77  \\
7  & coreutils  & gnubug-25023  & 0   & 10111  & 183  & -  & 3  & 119   \\
8  & coreutils  & gnubug-26545   & 0  & 286   & 101  & -  & 26  & 84 \\
9  & ffmpeg     & bugchrom-1404   & -  &-  & -  & -  & -  & - \\
10  & ffmpeg   & CVE-2017-9992  & -  & -  & -  & -  & -  & -   \\
11  & jasper  & CVE-2016-8691  & 127  & 2038 &  107 & 6  & 21   & 297 \\
12  & jasper  & CVE-2016-9557  & 5633 & 303  & 40  & - & 235  & 48    \\
13  & libarchive  & CVE-2016-5844  & 89 & 0 & 66  & 10 & 19  & 162  \\
14  & libjpeg  & CVE-2012-2806 & 0  & 0  & 136  & -  & 47  & 2159  \\
15  & libjpeg  & CVE-2017-15232 & 1 & 16 & 94  & 1  & 651  & 8 \\
16  & libjpeg  & CVE-2018-14498 & 903  & 1557 & 86  & -  & 30  & 73 \\
17  & libjpeg  & CVE-2018-19664  & 1208 & 137 & 36  & 4  & 791  & 3 \\
18  & libming  & CVE-2016-9264  & 1787 & 865 & 31  & - & 45  & 192  \\
19  & libming  & CVE-2018-8806  & 3165 & 959 & 45  & 1  & 5124  & 244 \\
20  & libming  & CVE-2018-8964  & 97 & 238 & 72  & 2 & 2011 & 68  \\
21  & libtiff  & bugzilla-2611  & 1204  & 28  & 85  & 1 & 677  & 13  \\
22  & libtiff  & bugzilla-2633  & 91   & 152  & 124  & 4   & 585  & 20  \\
23  & libtiff  & CVE-2016-10092 & 311  & 2985 & 183  & 1 & 1111  & 117   \\
24  & libtiff  & CVE-2016-10094 & 2503 & 106 & 504   & 1 & 480  & 103  \\
25  & libtiff  & CVE-2016-10272  & 206 & 931 & 184 & 1  & 984  & 99  \\
26  & libtiff  & CVE-2016-3186 & 768 & 1322  & 26 & 2  & 9  & 20   \\
27  & libtiff  & CVE-2016-5314  & 491  & 661 & 107  & 6 & 432  & 16  \\
28  & libtiff  & CVE-2016-5321  & 86  & 647 & 181 & - & 547  & 32 \\
29  & libtiff  & CVE-2016-9273  & 375 & 244 & 25  & -  & 388 & 16 \\
30  & libtiff  & CVE-2016-9532  & 2775  & 372 & 512 & 11  & 929 & 171  \\
31  & libtiff  & CVE-2017-5225  & 42 & 299  & 95 & 6 & 689  & 63  \\
32  & libtiff  & CVE-2017-7595  & 1  & 66 & 128  & -  & 396 & 12   \\
33  & libtiff  & CVE-2017-7599  & 3344  & 349  & 48 & - & 279  & 51  \\
34  & libtiff  & CVE-2017-7600  & 71 & 3042 & 41 & -  & 160 & 8  \\
35  & libtiff  & CVE-2017-7601  & 432 & 1338  & 121   & 1   & 231  & 19  \\
36  & libxml2  & CVE-2012-5134  & 99  & 115  & 363  & 4   & 236  & 258  \\
37  & libxml2  & CVE-2016-1838  & 14  & 142  & 254 & 1   & 1193 & 603 \\
38  & libxml2  & CVE-2016-1839  & 131  & 79  & 241  & 7  & 1024  & 297  \\
39  & libxml2  & CVE-2017-5969  & 0  & 143  & 94  & 1  & 329  & 120  \\
40  & potrace  & CVE-2013-7437  & 0  & 1213  & 21  & 2  & 6  & 16  \\
41  & zziplib  & CVE-2017-5974  & 433  & 185 & 59  & 7  & 34  & 53  \\
42  & zziplib  & CVE-2017-5975  & 1644  & 136  & 18  & 6  & 26  & 29 \\
43  & zziplib  & CVE-2017-5976  & 278  & 198  & 53  & -   & 22  & 9  \\                 
\hline
\hline 
\multicolumn{3}{|c|}{\textbf{Average} } & 777 & 2237 & 131 & - & 579 & 146 \\
\hline

\end{tabular}
\end{table}

%% file: related.tex
\section{Related Work}

In this section, we review related literature on greybox fuzzing and program repair.

\subsection{Grey Box Fuzzing}
Traditionally grey box fuzzing \cite{godefroid} has been used to generate tests with the goal of finding bugs in the program (e.g., \cite{aflfast}). These techniques observe program executions to identify undesired behaviors such as crashes or hangs while using coverage as feedback to guide the input generation. Although traditional usages of grey box fuzzing have been to discover program errors, it has also been used for fault localization~\cite{aurora, vulnloc} and program repair~\cite{learn2fix, difftgen, fix2fit, opad}. Fuzzing for fault localization generates large number of test-cases to be used with program analysis such as spectrum-based fault localization to identify the root cause of failure. In contrast, \toolName performs direct fix localization by generating and validating patches at runtime, which alleviates the necessity to generate a large number of test-cases, making it more efficient. In the context of program repair, fuzzing has been utilized to generate new inputs to identify overfitting patches. This line of work is most relevant to us, which generates test inputs, monitors the execution for the original program and patched program, and selects patches that fail on new inputs. DiffTGen~\cite{difftgen} identifies overfitting patches through test case generation, by using the fixed version of the program as the oracle. Opad~\cite{opad} uses memory safety properties, and Fix2Fit~\cite{fix2fit} utilizes security oracles from sanitizers to determine undesired behavior for the newly generated input. LEARN2FIX~\cite{learn2fix} generates new input and uses a learning model to predict undesired behavior.
All of these techniques rely on existing program repair techniques to generate patches, where fuzzing is used merely to identify and remove overfitting patches among an already constructed patch pool. The generated new inputs are not used by the run of the repair process itself but rather as a post-processing step. 

In contrast,
\toolName uses fuzzing to generate patches by searching over the space of program-edits and providing feedback for the search-based patch generation by simultaneously exploring the input space. Thus fuzzing is the search process to generate (and navigate) the patch space itself.

\toolName relies on two inherent oracles, namely crash-freedom and the buggy program itself. Security sanitizers are instrumented to detect program crashes, while the buggy program is used to determine the expected output for the new input. Similar to prior work~\cite{yu2017}, the assumption is in terms of expected behavior, where positive tests on the buggy and the patched program should behave similarly, while negative tests on the buggy and patched program should behave differently.

\subsection{Program Repair}
Automated program repair \cite{LPR19} was initially formulated as a co-evolutionary algorithm by Yao et al.~\cite{yao2008} where the idea is to evolve both the program and test cases for the program, to automatically fix bugs in the program. The proposed co-evolutionary approach requires a formal specification to define the expected behavior of the program, which limited the applicability in practice, since such formal specification is difficult to find. The first program repair technique applied to real-world software was GenProg~\cite{genprog}, which is an evolutionary algorithm that evolves the program with respect to a user-provided test-suite to search for test-adequate repairs. Relying on test-suite in contrast to a formal specification provided the scalability and applicability to program repair on real-world software. Several other evolutionary algorithm based repair techniques were proposed using different representations of the patch~\cite{LeGoues2012, par, arja, arja-e}. All these techniques evolve the program with respect to a fixed test-suite that generates patches potentially overfitting the given test suite~\cite{cure}. \toolName extends the idea of co-evolution proposed by Yao et al.~\cite{yao2008} but adapts the test-suite as in GrenProg~\cite{genprog}, relaxing the assumption of a formal specification. Existing evolutionary based repair techniques are limited in evolving only the program. In contrast, \toolName evolves both the program and the test-suite by co-exploration of both the program edit-space and input-space. 

Recent work on co-evolution for program repair has been shown to improve the quality of the generated patches by detecting and discarding overfitting patches by exploring the input-space with the goal of finding evidence to refute overfitting patches~\cite{cpr, fix2fit}. 
Fix2Fit~\cite{fix2fit} uses directed grey-box fuzzing with the aim of identifying inputs that crash on the patched program. CPR~\cite{cpr} uses concolic execution with the aid of a user-provided specification to generate new inputs to identify and discard overfitting patches. Both approaches explore the input-space to discard patches from an initialized set of patches, resulting in shrinking the initial patch-space. In comparison, \toolName evolves the patch-space rather than discarding a set of patches. \toolName mutates identified over-fitting patches to evolve such that they pass the updated fitness criteria (i.e., new test-cases). Thus the patch-pool is not only removed of over-fitting patches but also increased with new set of evolved patches. 

%% file: conclusion.tex
\section{Discussion}

Automated repair of security vulnerabilities in programs, is desirable. This is because of the large time lag (typically 60-150 days) between detection of vulnerabilities and fixing them. It is also estimated that we need a 50\% increase over today's staffing levels to achieve timely responses to vulnerabilities \cite{study-report}. While automated program repair provides a promising direction to produce fixes to vulnerabilities as they are detected, technically achieving this goal remains challenging for (at least) the following reasons. 
\begin{itemize}[leftmargin=*]
\item Test-based program repair produces fixes which work for the given test-suite but may fail for tests outside the given test-suite. 
\item For program vulnerability repair, very few tests are typically available --- often only one failing test in the form of an exploit.
\item Since few tests are available, there may be many patch candidates, hence validating the various patch candidates (often by fuzzing the fixed program to find crashes) turns out to be costly due to the recompilation time after patches are inserted into the vulnerable program. 
\end{itemize} 

In this work, we have provided a pragmatic solution to these challenges. Our core observation is that fuzzing is an extremely optimized biased random search process. Thus, apart from finding vulnerabilities in the input space, it can be re-purposed to search over the patch space to find fixes. This leads to program repair being accomplished as a collaboration between input-level fuzzer and patch-level fuzzer. Since the patch-level fuzzer accomplishes compilation free repair via binary rewriting, it is able to leverage the fast pace of fuzzing to efficiently explore large patch spaces. In fact by leveraging fuzzing as the core search engine in repair, we search over a very large search space which includes several candidate fix locations, as well as patch candidates in each location. We are thus not reliant on a developer provided fix location. Furthermore, and more importantly, even if our automatically generated fix is not directly used -- it yields specific insights such as possible fix locations which a developer can use to fix a vulnerability. Other by-products from our patching process such as the additional tests generated by the input level fuzzer can also aid in fixing; these artifacts can be useful for repair - as mentioned in recent surveys \cite{trustAPR}. Such aid in fixing vulnerabilities can significantly cut down the time lag between reporting and fixing vulnerabilities - thereby reducing the exposure of critical software systems to attacks. 

Overall, we do not view our work as {\em automated} program repair; instead it can come under the more realistic vision of {\em greater automation} in program repair. Moreover, since greybox fuzzing is often the technology of choice for bug detection, our approach brings bug repair closer to bug detection - in terms of design as well as implementation.

%% file: main.bbl

\begin{thebibliography}{40}


\ifx \showCODEN    \undefined \def \showCODEN     #1{\unskip}     \fi
\ifx \showDOI      \undefined \def \showDOI       #1{#1}\fi
\ifx \showISBNx    \undefined \def \showISBNx     #1{\unskip}     \fi
\ifx \showISBNxiii \undefined \def \showISBNxiii  #1{\unskip}     \fi
\ifx \showISSN     \undefined \def \showISSN      #1{\unskip}     \fi
\ifx \showLCCN     \undefined \def \showLCCN      #1{\unskip}     \fi
\ifx \shownote     \undefined \def \shownote      #1{#1}          \fi
\ifx \showarticletitle \undefined \def \showarticletitle #1{#1}   \fi
\ifx \showURL      \undefined \def \showURL       {\relax}        \fi
\providecommand\bibfield[2]{#2}
\providecommand\bibinfo[2]{#2}
\providecommand\natexlab[1]{#1}
\providecommand\showeprint[2][]{arXiv:#2}

\bibitem[afl(2020)]%
        {afl}
 \bibinfo{year}{2020}\natexlab{}.
\newblock \bibinfo{title}{Github Repository for American Fuzzy Lop}.
\newblock \bibinfo{howpublished}{\url{https://github.com/google/AFL}}.
\newblock
\newblock
\shownote{Accessed: 2022-10-12}.


\bibitem[dwa(2022)]%
        {dwarf}
 \bibinfo{year}{2022}\natexlab{}.
\newblock \bibinfo{title}{{DWARF Debugging Information Format Version 5}}.
\newblock \bibinfo{howpublished}{\url{https://dwarfstd.org/doc/DWARF5.pdf}}.
\newblock


\bibitem[gdb(2022)]%
        {gdb}
 \bibinfo{year}{2022}\natexlab{}.
\newblock \bibinfo{title}{The {GNU} {D}ebugger}.
\newblock
  \bibinfo{howpublished}{\url{https://sourceware.org/git/binutils-gdb.git}}.
\newblock


\bibitem[Arcuri and Yao(2008)]%
        {yao2008}
\bibfield{author}{\bibinfo{person}{Andrea Arcuri} {and} \bibinfo{person}{Xin
  Yao}.} \bibinfo{year}{2008}\natexlab{}.
\newblock \showarticletitle{A novel co-evolutionary approach to automatic
  software bug fixing}. In \bibinfo{booktitle}{\emph{2008 IEEE Congress on
  Evolutionary Computation (IEEE World Congress on Computational
  Intelligence)}}. \bibinfo{pages}{162--168}.
\newblock
\urldef\tempurl%
\url{https://doi.org/10.1109/CEC.2008.4630793}
\showDOI{\tempurl}


\bibitem[Bader et~al\mbox{.}(2019)]%
        {bader2019getafix}
\bibfield{author}{\bibinfo{person}{Johannes Bader}, \bibinfo{person}{Andrew
  Scott}, \bibinfo{person}{Michael Pradel}, {and} \bibinfo{person}{Satish
  Chandra}.} \bibinfo{year}{2019}\natexlab{}.
\newblock \showarticletitle{Getafix: Learning to fix bugs automatically}.
\newblock \bibinfo{journal}{\emph{Proceedings of the ACM on Programming
  Languages}} \bibinfo{volume}{3}, \bibinfo{number}{OOPSLA}
  (\bibinfo{year}{2019}), \bibinfo{pages}{1--27}.
\newblock


\bibitem[Beyer et~al\mbox{.}(2021)]%
        {dirk2021}
\bibfield{author}{\bibinfo{person}{Dirk Beyer}, \bibinfo{person}{Lars Grunske},
  \bibinfo{person}{Thomas Lemberger}, {and} \bibinfo{person}{Minxing Tang}.}
  \bibinfo{year}{2021}\natexlab{}.
\newblock \showarticletitle{Towards a Benchmark Set for Program Repair Based on
  Partial Fixes}.
\newblock \bibinfo{journal}{\emph{CoRR}}  \bibinfo{volume}{abs/2107.08038}
  (\bibinfo{year}{2021}).
\newblock
\showeprint[arXiv]{2107.08038}
\urldef\tempurl%
\url{https://arxiv.org/abs/2107.08038}
\showURL{%
\tempurl}


\bibitem[Blazytko et~al\mbox{.}(2020)]%
        {aurora}
\bibfield{author}{\bibinfo{person}{Tim Blazytko}, \bibinfo{person}{Moritz
  Schl{\"o}gel}, \bibinfo{person}{Cornelius Aschermann}, \bibinfo{person}{Ali
  Abbasi}, \bibinfo{person}{Joel Frank}, \bibinfo{person}{Simon W{\"o}rner},
  {and} \bibinfo{person}{Thorsten Holz}.} \bibinfo{year}{2020}\natexlab{}.
\newblock \showarticletitle{{AURORA}: Statistical Crash Analysis for Automated
  Root Cause Explanation}. In \bibinfo{booktitle}{\emph{29th USENIX Security
  Symposium (USENIX Security 20)}}. \bibinfo{publisher}{USENIX Association},
  \bibinfo{pages}{235--252}.
\newblock
\showISBNx{978-1-939133-17-5}
\urldef\tempurl%
\url{https://www.usenix.org/conference/usenixsecurity20/presentation/blazytko}
\showURL{%
\tempurl}


\bibitem[B{\"o}hme et~al\mbox{.}(2021)]%
        {fuzz-survey}
\bibfield{author}{\bibinfo{person}{Marcel B{\"o}hme}, \bibinfo{person}{Cristian
  Cadar}, {and} \bibinfo{person}{Abhik Roychoudhury}.}
  \bibinfo{year}{2021}\natexlab{}.
\newblock \showarticletitle{Fuzzing: Challenges and Reflections}.
\newblock \bibinfo{journal}{\emph{{IEEE} Software}} \bibinfo{volume}{38},
  \bibinfo{number}{3} (\bibinfo{year}{2021}).
\newblock


\bibitem[B{\"o}hme et~al\mbox{.}(2016)]%
        {aflfast}
\bibfield{author}{\bibinfo{person}{Marcel B{\"o}hme},
  \bibinfo{person}{Van-Thuan Pham}, {and} \bibinfo{person}{Abhik
  Roychoudhury}.} \bibinfo{year}{2016}\natexlab{}.
\newblock \showarticletitle{Coverage based Greybox Fuzzing as a Markov Chain}.
  In \bibinfo{booktitle}{\emph{Proceedings of the 2017 ACM SIGSAC Conference on
  Computer and Communications Security (CCS)}}.
\newblock


\bibitem[Böhme et~al\mbox{.}(2020)]%
        {learn2fix}
\bibfield{author}{\bibinfo{person}{Marcel Böhme}, \bibinfo{person}{Charaka
  Geethal}, {and} \bibinfo{person}{Van-Thuan Pham}.}
  \bibinfo{year}{2020}\natexlab{}.
\newblock \showarticletitle{Human-In-The-Loop Automatic Program Repair}. In
  \bibinfo{booktitle}{\emph{2020 IEEE 13th International Conference on Software
  Testing, Validation and Verification (ICST)}}. \bibinfo{pages}{274--285}.
\newblock
\urldef\tempurl%
\url{https://doi.org/10.1109/ICST46399.2020.00036}
\showDOI{\tempurl}


\bibitem[Duck et~al\mbox{.}(2020)]%
        {e9patch}
\bibfield{author}{\bibinfo{person}{Gregory~J. Duck}, \bibinfo{person}{Xiang
  Gao}, {and} \bibinfo{person}{Abhik Roychoudhury}.}
  \bibinfo{year}{2020}\natexlab{}.
\newblock \showarticletitle{Binary Rewriting without Control Flow Recovery}. In
  \bibinfo{booktitle}{\emph{Proceedings of the 41st ACM SIGPLAN Conference on
  Programming Language Design and Implementation}} (London, UK)
  \emph{(\bibinfo{series}{PLDI 2020})}. \bibinfo{publisher}{Association for
  Computing Machinery}, \bibinfo{address}{New York, NY, USA},
  \bibinfo{pages}{151–163}.
\newblock
\showISBNx{9781450376136}
\urldef\tempurl%
\url{https://doi.org/10.1145/3385412.3385972}
\showDOI{\tempurl}


\bibitem[Gao et~al\mbox{.}(2019)]%
        {fix2fit}
\bibfield{author}{\bibinfo{person}{Xiang Gao}, \bibinfo{person}{Sergey
  Mechtaev}, {and} \bibinfo{person}{Abhik Roychoudhury}.}
  \bibinfo{year}{2019}\natexlab{}.
\newblock \showarticletitle{Crash-Avoiding Program Repair}. In
  \bibinfo{booktitle}{\emph{Proceedings of the 28th ACM SIGSOFT International
  Symposium on Software Testing and Analysis}} (Beijing, China)
  \emph{(\bibinfo{series}{ISSTA 2019})}. \bibinfo{publisher}{Association for
  Computing Machinery}, \bibinfo{address}{New York, NY, USA},
  \bibinfo{pages}{8–18}.
\newblock
\showISBNx{9781450362245}
\urldef\tempurl%
\url{https://doi.org/10.1145/3293882.3330558}
\showDOI{\tempurl}


\bibitem[Godefroid(2020)]%
        {godefroid}
\bibfield{author}{\bibinfo{person}{Patrice Godefroid}.}
  \bibinfo{year}{2020}\natexlab{}.
\newblock \showarticletitle{Fuzzing: Hack, art and science}.
\newblock \bibinfo{journal}{\emph{Commun. ACM}} \bibinfo{volume}{63},
  \bibinfo{number}{2} (\bibinfo{year}{2020}), \bibinfo{pages}{70--76}.
\newblock


\bibitem[Goues et~al\mbox{.}(2019)]%
        {LPR19}
\bibfield{author}{\bibinfo{person}{Claire~Le Goues}, \bibinfo{person}{Michael
  Pradel}, {and} \bibinfo{person}{Abhik Roychoudhury}.}
  \bibinfo{year}{2019}\natexlab{}.
\newblock \showarticletitle{Automated Program Repair}.
\newblock \bibinfo{journal}{\emph{Commun. ACM}} \bibinfo{volume}{62},
  \bibinfo{number}{12} (\bibinfo{date}{Nov.} \bibinfo{year}{2019}),
  \bibinfo{pages}{56–65}.
\newblock


\bibitem[Huang et~al\mbox{.}(2019)]%
        {senx}
\bibfield{author}{\bibinfo{person}{Zhen Huang}, \bibinfo{person}{David Lie},
  \bibinfo{person}{Gang Tan}, {and} \bibinfo{person}{Trent Jaeger}.}
  \bibinfo{year}{2019}\natexlab{}.
\newblock \showarticletitle{Using Safety Properties to Generate Vulnerability
  Patches}. In \bibinfo{booktitle}{\emph{2019 IEEE Symposium on Security and
  Privacy (SP)}}. \bibinfo{pages}{539--554}.
\newblock
\urldef\tempurl%
\url{https://doi.org/10.1109/SP.2019.00071}
\showDOI{\tempurl}


\bibitem[Institute(2022)]%
        {study-report}
\bibfield{author}{\bibinfo{person}{Ponemon Institute}.}
  \bibinfo{year}{2022}\natexlab{}.
\newblock \bibinfo{title}{Costs and Consequences of Gaps in Vulnerability
  Response}.
\newblock
\newblock
\newblock
\shownote{\url{https://www.servicenow.com/lpayr/ponemon-vulnerability-survey.html}}.


\bibitem[Kim et~al\mbox{.}(2013)]%
        {par}
\bibfield{author}{\bibinfo{person}{Dongsun Kim}, \bibinfo{person}{Jaechang
  Nam}, \bibinfo{person}{Jaewoo Song}, {and} \bibinfo{person}{Sunghun Kim}.}
  \bibinfo{year}{2013}\natexlab{}.
\newblock \showarticletitle{Automatic Patch Generation Learned from
  Human-Written Patches}. In \bibinfo{booktitle}{\emph{Proceedings of the 2013
  International Conference on Software Engineering}} (San Francisco, CA, USA)
  \emph{(\bibinfo{series}{ICSE '13})}. \bibinfo{publisher}{IEEE Press},
  \bibinfo{pages}{802–811}.
\newblock
\showISBNx{9781467330763}


\bibitem[{Le Goues} et~al\mbox{.}(2012)]%
        {LeGoues2012}
\bibfield{author}{\bibinfo{person}{Claire {Le Goues}}, \bibinfo{person}{Michael
  Dewey-Vogt}, \bibinfo{person}{Stephanie Forrest}, {and}
  \bibinfo{person}{Westley Weimer}.} \bibinfo{year}{2012}\natexlab{}.
\newblock \showarticletitle{{A systematic study of automated program repair:
  Fixing 55 out of 105 bugs for {\$}8 each}}. In \bibinfo{booktitle}{\emph{2012
  34th International Conference on Software Engineering (ICSE)}}.
  \bibinfo{pages}{3--13}.
\newblock
\urldef\tempurl%
\url{https://doi.org/10.1109/ICSE.2012.6227211}
\showDOI{\tempurl}


\bibitem[Le~Goues et~al\mbox{.}(2012)]%
        {genprog12}
\bibfield{author}{\bibinfo{person}{Claire Le~Goues}, \bibinfo{person}{Thanh~Vu
  Nguyen}, \bibinfo{person}{Stephanie Forrest}, {and} \bibinfo{person}{Westley
  Weimer}.} \bibinfo{year}{2012}\natexlab{}.
\newblock \showarticletitle{GenProg: A Generic Method for Automatic Software
  Repair}.
\newblock \bibinfo{journal}{\emph{IEEE Transactions on Software Engineering}}
  \bibinfo{volume}{38}, \bibinfo{number}{1} (\bibinfo{year}{2012}),
  \bibinfo{pages}{54--72}.
\newblock
\urldef\tempurl%
\url{https://doi.org/10.1109/TSE.2011.104}
\showDOI{\tempurl}


\bibitem[Liu et~al\mbox{.}(2019a)]%
        {liu2019}
\bibfield{author}{\bibinfo{person}{Kui Liu}, \bibinfo{person}{Anil Koyuncu},
  \bibinfo{person}{Tegawendé~F. Bissyandé}, \bibinfo{person}{Dongsun Kim},
  \bibinfo{person}{Jacques Klein}, {and} \bibinfo{person}{Yves Le~Traon}.}
  \bibinfo{year}{2019}\natexlab{a}.
\newblock \showarticletitle{You Cannot Fix What You Cannot Find! An
  Investigation of Fault Localization Bias in Benchmarking Automated Program
  Repair Systems}. In \bibinfo{booktitle}{\emph{2019 12th IEEE Conference on
  Software Testing, Validation and Verification (ICST)}}.
  \bibinfo{pages}{102--113}.
\newblock
\urldef\tempurl%
\url{https://doi.org/10.1109/ICST.2019.00020}
\showDOI{\tempurl}


\bibitem[Liu et~al\mbox{.}(2019b)]%
        {tbar}
\bibfield{author}{\bibinfo{person}{Kui Liu}, \bibinfo{person}{Anil Koyuncu},
  \bibinfo{person}{Dongsun Kim}, {and} \bibinfo{person}{Tegawendé~F.
  Bissyandé}.} \bibinfo{year}{2019}\natexlab{b}.
\newblock \showarticletitle{TBar: Revisiting Template-based Automated Program
  Repair}. In \bibinfo{booktitle}{\emph{International Symposium on Software
  Testing and Analysis ({ISSTA})}}.
\newblock


\bibitem[LLVM(2022)]%
        {llvm}
\bibfield{author}{\bibinfo{person}{LLVM}.} \bibinfo{year}{2022}\natexlab{}.
\newblock \bibinfo{booktitle}{\emph{{The LLVM Compiler Infrastructure}}}.
\newblock
\urldef\tempurl%
\url{https://llvm.org}
\showURL{%
\tempurl}


\bibitem[Long and Rinard(2016)]%
        {prophet}
\bibfield{author}{\bibinfo{person}{Fan Long} {and} \bibinfo{person}{Martin
  Rinard}.} \bibinfo{year}{2016}\natexlab{}.
\newblock \showarticletitle{Automatic Patch Generation by Learning Correct
  Code}. In \bibinfo{booktitle}{\emph{Proceedings of the 43rd Annual ACM
  SIGPLAN-SIGACT Symposium on Principles of Programming Languages}} (St.
  Petersburg, FL, USA) \emph{(\bibinfo{series}{POPL '16})}.
  \bibinfo{publisher}{Association for Computing Machinery},
  \bibinfo{address}{New York, NY, USA}, \bibinfo{pages}{298–312}.
\newblock
\showISBNx{9781450335492}
\urldef\tempurl%
\url{https://doi.org/10.1145/2837614.2837617}
\showDOI{\tempurl}


\bibitem[Mechtaev et~al\mbox{.}(2018)]%
        {f1x}
\bibfield{author}{\bibinfo{person}{Sergey Mechtaev}, \bibinfo{person}{Xiang
  Gao}, \bibinfo{person}{Shin~Hwei Tan}, {and} \bibinfo{person}{Abhik
  Roychoudhury}.} \bibinfo{year}{2018}\natexlab{}.
\newblock \showarticletitle{Test-Equivalence Analysis for Automatic Patch
  Generation}.
\newblock \bibinfo{journal}{\emph{ACM Trans. Softw. Eng. Methodol.}}
  \bibinfo{volume}{27}, \bibinfo{number}{4}, Article \bibinfo{articleno}{15}
  (\bibinfo{date}{Oct.} \bibinfo{year}{2018}), \bibinfo{numpages}{37}~pages.
\newblock
\showISSN{1049-331X}
\urldef\tempurl%
\url{https://doi.org/10.1145/3241980}
\showDOI{\tempurl}


\bibitem[Mechtaev et~al\mbox{.}(2015)]%
        {directfix}
\bibfield{author}{\bibinfo{person}{Sergey Mechtaev}, \bibinfo{person}{Jooyong
  Yi}, {and} \bibinfo{person}{Abhik Roychoudhury}.}
  \bibinfo{year}{2015}\natexlab{}.
\newblock \showarticletitle{DirectFix: Looking for Simple Program Repairs}. In
  \bibinfo{booktitle}{\emph{ACM/IEEE International Conference on Software
  Engineering (ICSE)}}.
\newblock


\bibitem[Nguyen et~al\mbox{.}(2013)]%
        {semfix}
\bibfield{author}{\bibinfo{person}{H.D.T. Nguyen}, \bibinfo{person}{D. Qi},
  \bibinfo{person}{A. Roychoudhury}, {and} \bibinfo{person}{S. Chandra}.}
  \bibinfo{year}{2013}\natexlab{}.
\newblock \showarticletitle{SemFix: Program Repair via Semantic Analysis}. In
  \bibinfo{booktitle}{\emph{International Conference on Software Engineering}}.
\newblock


\bibitem[Noller et~al\mbox{.}(2021)]%
        {trustAPR}
\bibfield{author}{\bibinfo{person}{Yannic Noller}, \bibinfo{person}{Ridwan
  Shariffdeen}, \bibinfo{person}{Xiang Gao}, {and} \bibinfo{person}{Abhik
  Roychoudhury}.} \bibinfo{year}{2021}\natexlab{}.
\newblock \showarticletitle{Trust Enhancement Issues in Program Repair}.
\newblock \bibinfo{journal}{\emph{CoRR}}  \bibinfo{volume}{abs/2108.13064}
  (\bibinfo{year}{2021}).
\newblock
\showeprint[arXiv]{2108.13064}
\urldef\tempurl%
\url{https://arxiv.org/abs/2108.13064}
\showURL{%
\tempurl}


\bibitem[Serebryany et~al\mbox{.}(2012)]%
        {asan}
\bibfield{author}{\bibinfo{person}{Konstantin Serebryany},
  \bibinfo{person}{Derek Bruening}, \bibinfo{person}{Alexander Potapenko},
  {and} \bibinfo{person}{Dmitry Vyukov}.} \bibinfo{year}{2012}\natexlab{}.
\newblock \showarticletitle{AddressSanitizer: A Fast Address Sanity Checker}.
  In \bibinfo{booktitle}{\emph{Proceedings of the 2012 USENIX Conference on
  Annual Technical Conference}} (Boston, MA) \emph{(\bibinfo{series}{USENIX
  ATC'12})}. \bibinfo{publisher}{USENIX Association}, \bibinfo{address}{USA},
  \bibinfo{pages}{28}.
\newblock


\bibitem[Shariffdeen et~al\mbox{.}(2021)]%
        {cpr}
\bibfield{author}{\bibinfo{person}{Ridwan Shariffdeen}, \bibinfo{person}{Yannic
  Noller}, \bibinfo{person}{Lars Grunske}, {and} \bibinfo{person}{Abhik
  Roychoudhury}.} \bibinfo{year}{2021}\natexlab{}.
\newblock \showarticletitle{Concolic Program Repair}. In
  \bibinfo{booktitle}{\emph{Proceedings of the 42nd ACM SIGPLAN International
  Conference on Programming Language Design and Implementation}} (Virtual,
  Canada) \emph{(\bibinfo{series}{PLDI 2021})}. \bibinfo{publisher}{Association
  for Computing Machinery}, \bibinfo{address}{New York, NY, USA},
  \bibinfo{pages}{390–405}.
\newblock
\showISBNx{9781450383912}
\urldef\tempurl%
\url{https://doi.org/10.1145/3453483.3454051}
\showDOI{\tempurl}


\bibitem[Shen et~al\mbox{.}(2021)]%
        {vulnloc}
\bibfield{author}{\bibinfo{person}{Shiqi Shen}, \bibinfo{person}{Aashish
  Kolluri}, \bibinfo{person}{Zhen Dong}, \bibinfo{person}{Prateek Saxena},
  {and} \bibinfo{person}{Abhik Roychoudhury}.} \bibinfo{year}{2021}\natexlab{}.
\newblock \showarticletitle{Localizing Vulnerabilities Statistically From One
  Exploit}. In \bibinfo{booktitle}{\emph{Proceedings of the 2021 ACM Asia
  Conference on Computer and Communications Security}} (Virtual Event, Hong
  Kong) \emph{(\bibinfo{series}{ASIA CCS '21})}.
  \bibinfo{publisher}{Association for Computing Machinery},
  \bibinfo{address}{New York, NY, USA}, \bibinfo{pages}{537–549}.
\newblock
\showISBNx{9781450382878}
\urldef\tempurl%
\url{https://doi.org/10.1145/3433210.3437528}
\showDOI{\tempurl}


\bibitem[Smith et~al\mbox{.}(2015)]%
        {cure}
\bibfield{author}{\bibinfo{person}{Edward~K. Smith}, \bibinfo{person}{Earl~T.
  Barr}, \bibinfo{person}{Claire Le~Goues}, {and} \bibinfo{person}{Yuriy
  Brun}.} \bibinfo{year}{2015}\natexlab{}.
\newblock \showarticletitle{Is the Cure Worse than the Disease? Overfitting in
  Automated Program Repair} \emph{(\bibinfo{series}{ESEC/FSE 2015})}.
  \bibinfo{publisher}{Association for Computing Machinery},
  \bibinfo{address}{New York, NY, USA}, \bibinfo{pages}{532–543}.
\newblock
\showISBNx{9781450336758}
\urldef\tempurl%
\url{https://doi.org/10.1145/2786805.2786825}
\showDOI{\tempurl}


\bibitem[Team({[n.\,d.]})]%
        {ubsan}
\bibfield{author}{\bibinfo{person}{The~Clang Team}.}
  \bibinfo{year}{[n.\,d.]}\natexlab{}.
\newblock \bibinfo{title}{Undefined Behavior Sanitizer}.
\newblock
\newblock
\newblock
\shownote{\url{https://clang.llvm.org/docs/UndefinedBehaviorSanitizer.html}}.


\bibitem[Timperley et~al\mbox{.}({[n.\,d.]})]%
        {darjeeling}
\bibfield{author}{\bibinfo{person}{Christopher Timperley} {et~al\mbox{.}}}
  \bibinfo{year}{[n.\,d.]}\natexlab{}.
\newblock \bibinfo{title}{Darjeeling: language agnostic search-based repair
  tool}.
\newblock
\newblock
\newblock
\shownote{\url{https://github.com/squaresLab/Darjeeling}}.


\bibitem[Weimer et~al\mbox{.}(2009)]%
        {genprog}
\bibfield{author}{\bibinfo{person}{Westley Weimer}, \bibinfo{person}{ThanhVu
  Nguyen}, \bibinfo{person}{Claire~Le Goues}, {and} \bibinfo{person}{Stephanie
  Forrest}.} \bibinfo{year}{2009}\natexlab{}.
\newblock \showarticletitle{Automatically finding patches using genetic
  programming}. In \bibinfo{booktitle}{\emph{IEEE/ACM International Conference
  on Software Engineering ({ICSE})}}.
\newblock


\bibitem[Xin and Reiss(2017)]%
        {difftgen}
\bibfield{author}{\bibinfo{person}{Qi Xin} {and} \bibinfo{person}{Steven~P.
  Reiss}.} \bibinfo{year}{2017}\natexlab{}.
\newblock \showarticletitle{Identifying Test-Suite-Overfitted Patches through
  Test Case Generation} \emph{(\bibinfo{series}{ISSTA 2017})}.
  \bibinfo{publisher}{Association for Computing Machinery},
  \bibinfo{address}{New York, NY, USA}, \bibinfo{pages}{226–236}.
\newblock
\showISBNx{9781450350761}
\urldef\tempurl%
\url{https://doi.org/10.1145/3092703.3092718}
\showDOI{\tempurl}


\bibitem[Xiong et~al\mbox{.}(2018)]%
        {Xiong2018_PatchCorrectness}
\bibfield{author}{\bibinfo{person}{Yingfei Xiong}, \bibinfo{person}{Xinyuan
  Liu}, \bibinfo{person}{Muhan Zeng}, \bibinfo{person}{Lu Zhang}, {and}
  \bibinfo{person}{Gang Huang}.} \bibinfo{year}{2018}\natexlab{}.
\newblock \showarticletitle{Identifying Patch Correctness in Test-Based Program
  Repair}. In \bibinfo{booktitle}{\emph{Proceedings of the 40th International
  Conference on Software Engineering}} (Gothenburg, Sweden)
  \emph{(\bibinfo{series}{ICSE '18})}. \bibinfo{publisher}{Association for
  Computing Machinery}, \bibinfo{address}{New York, NY, USA},
  \bibinfo{pages}{789–799}.
\newblock
\showISBNx{9781450356381}
\urldef\tempurl%
\url{https://doi.org/10.1145/3180155.3180182}
\showDOI{\tempurl}


\bibitem[Yang et~al\mbox{.}(2017)]%
        {opad}
\bibfield{author}{\bibinfo{person}{Jinqiu Yang}, \bibinfo{person}{Alexey
  Zhikhartsev}, \bibinfo{person}{Yuefei Liu}, {and} \bibinfo{person}{Lin Tan}.}
  \bibinfo{year}{2017}\natexlab{}.
\newblock \showarticletitle{Better Test Cases for Better Automated Program
  Repair}. In \bibinfo{booktitle}{\emph{Proceedings of the 2017 11th Joint
  Meeting on Foundations of Software Engineering}} (Paderborn, Germany)
  \emph{(\bibinfo{series}{ESEC/FSE 2017})}. \bibinfo{publisher}{Association for
  Computing Machinery}, \bibinfo{address}{New York, NY, USA},
  \bibinfo{pages}{831–841}.
\newblock
\showISBNx{9781450351058}
\urldef\tempurl%
\url{https://doi.org/10.1145/3106237.3106274}
\showDOI{\tempurl}


\bibitem[Yu et~al\mbox{.}(2017)]%
        {yu2017}
\bibfield{author}{\bibinfo{person}{Zhongxing Yu}, \bibinfo{person}{Matias
  Martinez}, \bibinfo{person}{Benjamin Danglot}, \bibinfo{person}{Thomas
  Durieux}, {and} \bibinfo{person}{Martin Monperrus}.}
  \bibinfo{year}{2017}\natexlab{}.
\newblock \showarticletitle{Test Case Generation for Program Repair: {A} Study
  of Feasibility and Effectiveness}.
\newblock \bibinfo{journal}{\emph{CoRR}}  \bibinfo{volume}{abs/1703.00198}
  (\bibinfo{year}{2017}).
\newblock
\showeprint[arXiv]{1703.00198}
\urldef\tempurl%
\url{http://arxiv.org/abs/1703.00198}
\showURL{%
\tempurl}


\bibitem[Yuan and Banzhaf(2020a)]%
        {arja}
\bibfield{author}{\bibinfo{person}{Yuan Yuan} {and} \bibinfo{person}{Wolfgang
  Banzhaf}.} \bibinfo{year}{2020}\natexlab{a}.
\newblock \showarticletitle{ARJA: Automated Repair of Java Programs via
  Multi-Objective Genetic Programming}.
\newblock \bibinfo{journal}{\emph{IEEE Transactions on Software Engineering}}
  \bibinfo{volume}{46}, \bibinfo{number}{10} (\bibinfo{year}{2020}),
  \bibinfo{pages}{1040--1067}.
\newblock
\urldef\tempurl%
\url{https://doi.org/10.1109/TSE.2018.2874648}
\showDOI{\tempurl}


\bibitem[Yuan and Banzhaf(2020b)]%
        {arja-e}
\bibfield{author}{\bibinfo{person}{Yuan Yuan} {and} \bibinfo{person}{Wolfgang
  Banzhaf}.} \bibinfo{year}{2020}\natexlab{b}.
\newblock \showarticletitle{Toward Better Evolutionary Program Repair: An
  Integrated Approach}.
\newblock \bibinfo{journal}{\emph{ACM Trans. Softw. Eng. Methodol.}}
  \bibinfo{volume}{29}, \bibinfo{number}{1}, Article \bibinfo{articleno}{5}
  (\bibinfo{date}{jan} \bibinfo{year}{2020}), \bibinfo{numpages}{53}~pages.
\newblock
\showISSN{1049-331X}
\urldef\tempurl%
\url{https://doi.org/10.1145/3360004}
\showDOI{\tempurl}


\end{thebibliography}
